\documentclass[10pt,preprint,numbers,nocopyrightspace]{sigplanconf}
 
\usepackage{url}
\usepackage{stmaryrd}
\usepackage{galois}
\usepackage{proof,latexsym,amsmath,amssymb,amsfonts,mathrsfs,amsthm}
\usepackage{array}
\usepackage{pgf}
\usepackage{tikz}
\usetikzlibrary{arrows,automata,snakes}
\usepackage{pgfplots}
\usetikzlibrary{patterns}
\usepackage{fancyvrb} 
\usepackage{ifthen}
\usepackage{multirow} 
\usepackage{algorithm}
\usepackage{algpseudocode}
\usepackage{flushend} % balance the last page
\usepackage{qtree}
\usepackage{natbib}

\tikzset{
    vertex/.style = {
        circle,
        fill            = black,
        outer sep = 2pt,
        inner sep = 1pt,
    }
}

\newcommand{\needrevision}[1]{{\color{blue}{\bf Need revision}{}~}}

\newcommand{\longversion}[1]{#1}

 \newcommand{\hide}[1]{}

\newcommand{\mbsf}[1]{\mbox{\sf #1}}

\newcommand{\mca}{\mathcal{A}}

\newcommand{\Def}{\mbsf{D}}
\newcommand{\Use}{\mbsf{U}}

\newcommand{\mbb}{\mathbb{B}}
\newcommand{\mbc}{\mathbb{C}}
\newcommand{\mbd}{\mathbb{D}}

\newcommand{\mbj}{\mathbb{J}}

\newcommand{\mbo}{\mathbb{O}}
\newcommand{\mbp}{\mathbb{P}}
\newcommand{\mbq}{\mathbb{Q}}

\newcommand{\mbs}{\mathbb{S}}

\newcommand{\mbv}{\mathbb{V}}

\newcommand{\cfgto}{\hookrightarrow}
\newcommand{\dugto}{\leadsto}

\newcommand{\zap}[1]{}

\newcommand{\node}{n}

\newcommand{\myset}[1]{\{ #1 \}}

\newcommand{\Var}{\mathsf{Var}}

\newcommand{\aF}{\abst{F}}

\newcommand{\defusea}{\rightsquigarrow}

\newcommand{\defuse}[1]{\stackrel{#1}{\defusea}}

\newcommand{\power}[1]{\wp(#1)}

\newcommand{\myland}{\;\land\;}

\newtheorem{example}{Example}

\newcommand{\todoc}[2]{{\textcolor{#1} {\textbf{[[#2]]}}}}

\newcommand{\todoblue}[1]{\todoc{blue}{#1}}

\newcommand{\hy}[1]{\todoblue{HY: #1}}

\newcommand{\argmax}{\operatornamewithlimits{argmax}}

\renewcommand{\vec}[1]{\mathbf{#1}}
\newcommand{\component}{\mbj}
\newcommand{\idx}{j}
\newcommand{\pgm}{\mbp}
\renewcommand{\aF}{{F}}
\newcommand{\Query}{\mbq}

\newcommand{\classifier}{{\mathcal{C}}}
\newcommand{\strategy}{{\mathcal{H}}}

\newcommand{\refeq}[1]{(\ref{#1})}

\newcommand{\Pgm}{\mbp}

\newcommand{\reducer}{{\sf reduce}}

\newcommand{\binop}{\oplus}
\newcommand{\compare}{\prec}

\newcommand{\StaticAnalyzer}{F}

\newcommand{\fpred}{{\sf match}}

\newcommand{\reducecond}{\phi}

\newcommand{\assumecmd}{{\it assume}}
\newcommand{\assertcmd}{{\it assert}}
\newcommand{\alloccmd}{{\it alloc}}

\newcommand{\feat}{\pi}
\newcommand{\Feat}{\Pi}

\newcommand{\query}{\mbq}

\newcommand{\dep}{{\sf req}}

\newcommand{\pred}{{\sf pred}}

\newcommand{\GenerateFP}{{\sf gen\_fp}}
\newcommand{\GenerateDFG}{{\sf build\_dfg}}
\newcommand{\Learn}{{\sf learn}}
\newcommand{\evaluate}{{\sf evaluate}}

\newcommand{\positive}{\vec{P}_{\it pos}}

\newcommand{\positiveminimal}{\vec{P}_{\it feat}}

\newcommand{\param}{\mca}

\newcommand{\Oct}{\mbo}

\newcommand{\PODom}{\mbo_{\Gamma}}

\newcommand{\Ours}{\mbox{Ours}}
\newcommand{\FIi}{\mbox{\sc FIi}}
\newcommand{\FSi}{\mbox{\sc FSi}}
\newcommand{\FIp}{\mbox{\sc FIp}}
\newcommand{\FSp}{\mbox{\sc FSp}}
\newcommand{\Impact}{\mbox{\sc IMPCT}}

\title{Automatically Generating Features for Learning Program Analysis
  Heuristics}

 \authorinfo{Kwonsoo Chae \and Hakjoo Oh}{Korea University}{\{kchae,hakjoo\_oh\}@korea.ac.kr}
 \authorinfo{Kihong Heo}{Seoul National University}{khheo@ropas.snu.ac.kr}
 \authorinfo{Hongseok Yang}{University of Oxford}{hongseok.yang@cs.ox.ac.uk}
\begin{document}
\toappear{}

\maketitle{}

\begin{abstract}
  We present a technique for automatically generating features for
  data-driven program analyses.  Recently data-driven approaches for
  building a program analysis have been proposed, which mine existing
  codebases and automatically learn heuristics for finding a
  cost-effective abstraction for a given analysis task. Such
  approaches reduce the burden of the analysis designers, but they do
  not remove it completely; they still leave the highly nontrivial
  task of designing so called features to the hands of the designers.
  Our technique automates this feature design process.  The idea is to
  use programs as features after reducing and abstracting them.  Our
  technique goes through selected program-query pairs in codebases,
  and it reduces and abstracts the program in each pair to a few lines
  of code, while ensuring that the analysis behaves similarly for the
  original and the new programs with respect to the query.  Each
  reduced program serves as a boolean feature for program-query
  pairs. This feature evaluates to true for a given program-query pair
  when (as a program) it is included in the program part of the pair.
  We have implemented our approach for three real-world program
  analyses. Our experimental evaluation shows that these analyses with
  automatically-generated features perform comparably to those with
  manually crafted features.

\end{abstract}

%%% Local Variables:
%%% mode: latex
%%% TeX-master: "paper"
%%% End:

% \category{F.3.2}{Semantics of Programming Languages}{Program Analysis}
%  \keywords
% Adaptive Program Analysis, Machine Learning, Automatic Feature Construction

\section{Introduction}

In an ideal world, a static program analysis adapts to a given task
automatically, so that it uses expensive techniques for improving
analysis precision only when those techniques are absolutely
necessary. In a real world, however, most static analyses are not
capable of doing such automatic adaptation. Instead, they rely on
fixed manually-designed heuristics for deciding when these
precision-improving but costly techniques should be applied.  These
heuristics are usually suboptimal and brittle. More importantly, they
are the outcomes of a substantial amount of laborious engineering
efforts of analysis designers.

Addressing these concerns with manually-designed heuristics has been
the goal of a large body of research in the program-analysis
community~\cite{slam,clarke:jacm03,blast:popl04,rybal1,gulavani:tacas08,XinMGNY14,XinNY13,Smaragdakis:2014,OhLHYY14,NaikYCS12,GuptaMR13,Kastrinis2013}.
Recently researchers started to explore data-driven approaches, where
a static analysis uses a parameterized heuristic and the parameter
values that maximize the analysis performance are learned
automatically from existing codebases via machine learning
techniques~\cite{OhYaYi15,Grigore2016,HeoOhYa16,ChaJeOh16}; the
learned heuristic is then used for analyzing previously unseen
programs. The approaches have been used to generate various
cost-effective analysis heuristics automatically, for example, for
controlling the degree of flow or context-sensitivity~\cite{OhYaYi15},
or determining where to apply relational analysis~\cite{HeoOhYa16}, or
deciding the threshold values of widening operators~\cite{ChaJeOh16}.

However, these data-driven approaches have one serious drawback. Their
successes crucially depend on the qualities of so called features,
which convert analysis inputs, such as programs and queries, to the
kind of inputs that machine learning techniques understand. Designing
a right set of features requires a nontrivial amount of knowledge and
efforts of domain experts.  Furthermore, the features designed for one
analysis do not usually generalize to others. For example,
in~\cite{OhYaYi15}, a total of 45 features were manually designed for
controlling flow-sensitivity, but a new set of 38 features were needed
for controlling context-sensitivity. This manual task of crafting
features is a major impediment to the widespread adoption of
data-driven approaches in practice, as in other applications of
machine learning techniques.

In this paper, we present a technique for automatically generating
features for data-driven static program analyses. From existing
codebases, a static analysis with our technique learns not only an
analysis heuristic but also features necessary to learn the heuristic
itself. In the first phase of this learning process, a set of features
appropriate for a given analysis task is generated from given
codebases. The next phase uses the generated features and learns an
analysis heuristic from the codebases.  Our technique is underpinned
by two key ideas.  The first idea is to run a generic program reducer
(e.g., C-Reduce~\cite{creduce}) on the codebases with a static
analysis as a subroutine, and to synthesize automatically {\em feature
  programs}, small pieces of code that minimally describe when it is
worth increasing the precision of the analysis.  Intuitively these
feature programs capture programming patterns whose analysis results
benefit greatly from the increased precision of the analysis.  The
second idea is to generalize these feature programs and represent them
by abstract data-flow graphs. Such a graph becomes a boolean predicate
on program slices, which holds for a slice when the graph is included
in it.  We incorporate these ideas into a general framework that is
applicable to various parametric static analyses.

We show the effectiveness and generality of our technique by applying
it to three static analyses for the C programming language: partially
flow-sensitive interval and pointer analyses, and partial Octagon
analysis. Our technique successfully generated features relevant to
each analysis, which were then used for learning an effective analysis
heuristic. The experimental results show that the heuristics learned
with automatically-generated features have performance comparable to
those with hand-crafted features by analysis experts.

\paragraph{Contributions}
We summarize our contributions below.
\begin{itemize}
\item We present a framework for automatically generating features for
  learning analysis heuristics. The framework is general enough to be
  used for various kinds of analyses for the C programming language
  such as interval, pointer, and Octagon analyses.

\item We present a novel method that uses a program reducer 
  for generating good feature programs, which capture important
  behaviors of static analysis. 
  
\item We introduce the notion of abstract data-flow graphs and show
  how they can serve as generic features for data-driven static analyses.

\item We provide extensive experimental evaluations with three different kinds of
  static analyses.
\end{itemize}

\longversion{
\paragraph{Outline}

We informally describe our approach in Section~\ref{sec:overview}.
The formal counterparts of this description take up the next
four sections: Section~\ref{sec:staticanalysis} for the definition of
 parametric static analyses considered in the paper,
Section~\ref{sec:learning} for an algorithm that learns heuristics
for choosing appropriate parameter values for a given analysis task,
Section~\ref{sec:feat-generation} for our technique for automatically generating
features, and  Section~\ref{sec:instances} for instance analyses designed
according to our approach. In Section \ref{sec:experiments}, we report
the findings of the experimental evaluation of our approach. In Sections~\ref{sec:related} 
and \ref{sec:conclusion}, we explain the relationship between our approach and
other prior works, and finish the paper with concluding remarks. 

}

%%% Local Variables:
%%% mode: latex
%%% TeX-master: "paper"
%%% End:

\section{Overview}
\label{sec:overview}

We illustrate our approach using its
instantiation with a partially flow-sensitive interval analysis. 

%\subsection{A Partially Flow-Sensitive Interval Analysis}

Our interval analysis is query-based and partially flow-sensitive.
It uses a classifier $\classifier$ that 
predicts, for each query in a given program, whether
flow-sensitivity is crucial for proving the query:
the query can be proved with flow-sensitivity but not without it. 
If the prediction is
positive, the analysis applies flow-sensitivity (FS) to the program
variables that may influence the query: it computes the data-flow slice of
the query and tracks the variables in the slice flow-sensitively. 
On the other hand, if the prediction is negative,
the analysis applies flow-insensitivity (FI) to the variables on which the 
query depends.

For example, consider the following program:
\begin{Verbatim}[frame=single,numbers=left]
x = 0; y = 0; z = input(); w = 0;
y = x; y++;
assert (y > 0);   // Query 1
assert (z > 0);   // Query 2
assert (w == 0);  // Query 3
\end{Verbatim}
The first query needs FS to prove, and the second is
impossible to prove because the value of {\tt z} comes from the external input.
The last query is easily proved even with FI. Ideally,
we want the classifier to give positive prediction only to the first query,
so that our analysis keeps flow-sensitive results 
only for the variables {\tt x}
and {\tt y}, on which the first query depends, and analyzes other
variables flow-insensitively. That is, we want the analysis 
to compute the following result:
\begin{center}
\begin{tabular}{|c|l|c|} \hline
\multicolumn{2}{|c|}{flow-sensitive result} & flow-insensitive result\\ \cline{1-3}
line & \multicolumn{1}{c|}{abstract state} & abstract state \\ \hline
1 & $\{ x \mapsto [0,0], y \mapsto [0,0] \}$&  \\ 
2 & $\{ x \mapsto [0,0], y\mapsto [1,1]\}$ & %$ \{ z \mapsto
                                             % [-\infty,+\infty]$
                                              \\ 
3 & $\{ x \mapsto [0,0], y\mapsto [1,1]\}$ &$\{ z \mapsto \top, w \mapsto [0,0]\}$ \\
4 & $\{ x \mapsto [0,0], y\mapsto [1,1]\}$ & \\ 
5 & $\{ x \mapsto [0,0], y \mapsto [1,1]\}$ & \\   \hline
\end{tabular}
\end{center}
% \kwonsoo{z should have [-oo,+oo], not [0,0]}
Note that for {\tt x} and {\tt y},
the result keeps a separate abstract state at each program point, but 
for the other variables {\tt z} and {\tt w}, it has 
just one abstract state for all program points.

\begin{figure*}[t]
\begin{center}

\begin{tabular}{c@{\qquad\qquad\qquad}c@{\qquad\qquad}c}
\begin{minipage}{1.8in}
\begin{Verbatim}[numbers=left]
a = 0; 
while (1) {
 b = unknown();
 if (a > b)
   if (a < 3)
     assert (a < 5); 
 a++;
}
\end{Verbatim}
\end{minipage}
&
\begin{minipage}{1.3in}
\begin{Verbatim}[numbers=left]
a = 0; 
while (1) {
 if (a < 3)
   assert (a < 5);
 a++;
}
\end{Verbatim}
\end{minipage}
&
\begin{minipage}{1.5in}
\begin{tikzpicture}
  \node[draw] (Assign1) at (0,0) {${\sf x} := {\sf c}$};
  \node[draw] (Assume) at (1.5,0)
  {${\sf x} < {\sf c}$};
  \node[draw,fill=white,text=black] (Query) at (3.2,0) {$Q({\sf
      x} < {\sf c})$};
  \node[draw] (Assign2) at (1.5,1) {${\sf x} := {\sf x } + {\sf c}$};  
  \draw[->] (Assign1) to (Assume);
  \draw[->] (Assume) to (Query);
  \draw[->] (Assume) to[in=-150,out=150] (Assign2);
  \draw[->] (Assign2) to[in=30,out=-30] (Assume);
\end{tikzpicture}
\end{minipage}
\\
(a) Original program & (b) Feature program & (c) Abstract data-flow
                                             graph (Feature)
\vspace*{-2mm}
\end{tabular}
\end{center} 
\caption{Example program and feature. % From the original program (a),
%   our approach first reduces it and generates the feature program (b),
% and then abstracts the feature program into the abstract data-flow
% graph (c).
}
\vspace*{-2mm}
\label{fig:overview}
\end{figure*}
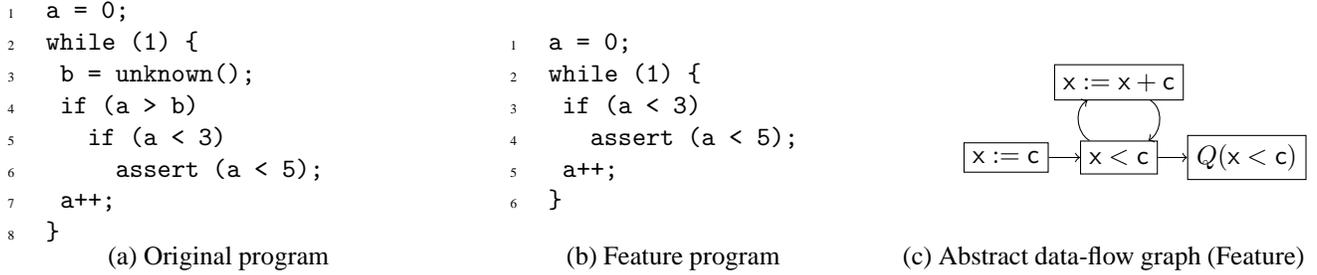

\subsection{Learning a Classifier}
The performance of the analysis crucially 
depends on the quality of its classifier $\classifier$. 
Instead of designing the classifier manually, we 
learn it from given codebases automatically. Let us illustrate
this learning process with a simple codebase of just one program $P$.

The input to the classifier learning is the collection
$\myset{(q^{(i)},b^{(i)})}_{i=1}^n$ of queries in $P$ labeled with
values $0$ and $1$. The label $b^{(i)}$ indicates whether the corresponding
query $q^{(i)}$ can be proved with FS but not with FI. These labeled
data are automatically generated by analyzing the 
codebase $\{P\}$ and identifying the queries that are proved with FS
but not with FI.

Given such data $\myset{(q^{(i)},b^{(i)})}_{i=1}^n$, 
we learn a classifier $\classifier$ in two steps.
First, we represent each query $q^{(i)}$ by a {\em feature vector}, 
which encodes essential properties of the query $q^{(i)}$
in the program $P$ and helps learning algorithms to achieve
good generalization. Formally, we transform 
the original data $\myset{(q^{(i)},b^{(i)})}_{i=1}^n$ to 
$\myset{(v^{(i)},b^{(i)})}_{i=1}^n$, where $v^{(i)} \in \mbb^k = \{0,1\}^k$ is a binary
feature vector of query $q^{(i)}$. 
The dimension $k$ of feature vectors
denotes the number of features. Second, to this transformed
data set $\myset{(v^{(i)},b^{(i)})}_{i=1}^n$, 
we apply an off-the-shelf
classification algorithm (such as decision tree)
and learn a classifier $\classifier: \mbb^k \to \mbb$, 
which takes a feature vector of a query and makes a prediction.

The success of this learning process relies heavily on
how we convert queries to feature vectors. If the feature vector of
a query ignores important information about the query for
prediction, learning a good classifier is impossible irrespective of
learning algorithms used. In previous work~\cite{OhYaYi15,HeoOhYa16,ChaJeOh16},
this feature engineering is done manually by analysis designers. For a specific
static analysis, they defined a set of \emph{features}
and used them to convert a query to a feature vector.
% Hakjoo: this is minor
% \kwonsoo{Feedback from labmates: In OOPSLA'15 and APLAS'16,
% things that are converted to feature vectors are program parts (variables in OOPSLA'15,
% constants in APLAS'16, etc.,), not queries.} 
But as in other applications of machine
 learning, this feature engineering requires
considerable domain expertise and engineering efforts. 
Our goal is to automatically generate high-quality features 
for this program-analysis application.

\subsection{Automatic Feature Generation}
\label{sec:overview-feature-generation}

We convert queries to feature vectors using a set
of features  $\Feat = \myset{\feat_1,\dots,\feat_k}$ and a 
procedure $\fpred$.  A feature $\feat_i$ encodes a property
about queries. The $\fpred$ procedure takes 
a feature $\feat$, a query $q_0$ and a program $P_0$ containing the query,
and checks whether the slice of $P_0$ that may affect $q_0$ satisfies
the property encoded by $\feat$. If so, it returns $1$, and otherwise, $0$. 
Using $\Feat$ and $\fpred$, we transform every query $q$ in the program $P$ 
of our codebase into a feature vector $v$:
\vspace*{-1mm}
\[
v = \langle \fpred(\feat_1,q,P), \dots, \fpred(\feat_k,q,P) \rangle. 
\vspace*{-0.4mm}
\]

\subsubsection{Feature Generation}
\label{sec:auto-feat-gen}

%In machine learning, a good feature should be both selective and
%invariant: the feature must be selective to the important aspects for
%discrimination but at the same time it must be also invariant and
%robust to the irrelevant aspects for generalization~\cite{DeepLearning}.  In our case, a
%good feature should describe a key property of flow-sensitivity, which
%distinguishes FS-provable queries from unprovable ones, and should not
%describe other irrelevant properties to enhance invariance.  We
%automatically generate such features by running a program reducer on
%the codebase and representing the results by abstract data flow
%graphs.

The unique aspect of our approach lies in our technique for generating
the feature set $\Pi$ from given codebases automatically.
\longversion{\footnote{In
our implementation, we partition the codebases to two groups. Programs
in the first group are used for feature generation and learning and
those in the other group are used for cross validation.}}
Two ideas make this automatic generation possible.

\paragraph{Generating Feature Programs Using a Reducer}

The first idea is to use a generic program reducer.
A reducer (e.g., C-Reduce~\cite{creduce}) takes a program 
and a predicate, and iteratively removes parts of the program 
as long as the predicate holds. 
%The result is a minimal program with the desired property.

We use a reducer to generate a collection of small code snippets that 
describe % typical \kwonsoo{Should we say 'typical' here? Maybe it is not what we want
% to say that the reducer captures only typical cases? I'm not sure.}
cases where the analysis can prove
a query with FS but not with FI. We first collect a set of queries
in codebases that require FS to prove. % can be proved by the analysis with FS,
% but not without it.
Then, for every query in the set,
we run the reducer on the program containing the query under
the predicate that the query in the reduced program continues to be FS-provable
but FI-unprovable. The reducer removes all the 
parts from the program that are irrelevant to the FS-provability
and the FI-unprovability of the query, leading to a
{\em feature program}. 

For example, consider the example program in Figure~\ref{fig:overview}(a).  
% \begin{equation}
% \begin{minipage}{2.1in}
% \begin{Verbatim}[numbers=left]
% a = 0; b = 0;
% while (1) {
%  b = unknown();
%  if (a > b)
%    if (a < 3)
%      assert (a < 5);  // Query
%  a++;
% }
% \end{Verbatim}
% \end{minipage}
% \end{equation}
The assertion at line 6 can be proved by the flow-sensitive interval
analysis but not by the flow-insensitive one; with FS, the
value of {\tt a} is restricted to the interval $[0,3]$ because of the
condition at line 5. With FI, {\tt a} has 
$[0,+\infty]$ at all program points. We reduce this program as
long as the flow-sensitive analysis proves the assertion while the
flow-insensitive one does not, resulting in the 
program in Figure~\ref{fig:overview}(b).
% \begin{equation}
% \begin{minipage}{2.1in}
% \begin{Verbatim}[numbers=left]
% a = 0; 
% while (1) {
%   if (a < 3)
%    assert (a < 5);  // Query
%  a++;
% }
% \end{Verbatim}
% \end{minipage}
% \end{equation}
Note that the reduced program only contains the key reasons
(i.e., loop and {\tt if (a < 3)}) for why FS works.
%but loses the parts (Figure
%\ref{fig:overview}(a)) that are irrelevant to the query. 
For example, the command {\tt if (a > b)} is removed because even without it,
the flow-sensitive analysis proves the query.
Running the reducer this way automatically removes these irrelevant parts of
the original program.

In experiments, we
used C-Reduce~\cite{creduce}, which has been used for
generating small test cases that trigger compiler bugs. The original program in
Figure \ref{fig:overview}(a) is too simplistic and does not
fully reflect the amount of slicing done by C-Reduce for real
programs. In our experiments, we found that C-Reduce is able
to transform programs with $>$1KLOC to those with just 5--10 lines,
similar to the one in Figure~\ref{fig:overview}(b). 

%In Section~\ref{sec:feat-generation}, we describe this
%feature generation technique in detail. 
% In particular, because simply using the
% reducer as an off-the-shelf tool is unlikely to preserve the original
% features, we need techniques to make original features survive after
% reduction (Section \ref{sec:preserving-features}).

\paragraph{Representing Feature Programs by Abstract Data-flow Graphs}
%\paragraph{Abstract Data-Flow Graphs}

The second idea is to represent the feature programs by abstract 
data-flow graphs. We build graphs that describe the data flows of the
feature programs. Then, we abstract individual atomic commands
in the graphs, for instance, by replacing some constants and variables
with the same fixed symbols ${\tt c}$ and ${\tt x}$, respectively.
The built graphs form the collection of features $\Feat$.

For example, the feature program in Figure \ref{fig:overview}(b) is 
represented by the graph in Figure \ref{fig:overview}(c).
The graph captures the data flows of the feature program that
influence the query. At the same time, the graph generalizes
the program by abstracting its atomic commands. All the variables
are replaced by the same symbol {\tt x}, and all
integers by {\tt c}, which in particular makes
the conditions {\tt a < 3} and {\tt a < 5} the same abstract
condition {\sf x $<$ c}.

How much should we abstract commands of the feature program? 
The answer depends on a static analysis.
If we abstract commands aggressively, this would introduce a strong
inductive bias, so that the algorithm for learning a classifier
might have hard time for finding a good classifier for given codebases
but would require fewer data for generalization. Otherwise, the
opposite situation would occur. Our technique considers multiple abstraction
levels, and automatically picks one to a given static analysis
using the combination of searching and cross-validation 
(Section~\ref{sec:feat-generation}). 

%  Each code snippets discovered from the codebase
% is transformed to a data flow graph. We define the feature set $\Feat$
% to be the set of all such graphs generated from the given codebase.

\subsubsection{Matching Algorithm}
\label{sec:feat-match}

By using the technique explained so far, 
we generate an abstract data-flow graph for each
FS-provable but FI-unprovable query in given codebases. These
graphs form the set of features, $\Feat = \myset{\feat_1,\dots,\feat_k}$.

The $\fpred$ procedure takes a feature (i.e., abstract data-flow graph) $\feat_i \in \Feat$, 
a query $q_0$, and a program $P_0$ containing $q_0$. Then,
it checks whether
the slice of $P_0$ that may affect $q_0$ includes a piece of code described by $\feat_i$.
Consider the query in the original program in Figure
\ref{fig:overview}(a) and the feature $\pi$ in Figure \ref{fig:overview}(c).
% $$
% \begin{minipage}{2.1in}
% \begin{Verbatim}[numbers=left]
% a = 0; b = 0;
% while (1) {
%  b = unknown();
%  if (a > b)
%    if (a < 3)
%      assert (a < 5);  // Query
%  a++;
% }
% \end{Verbatim}
% \end{minipage}
% $$
% We would like to represent the query by a feature vector. The feature
% vector $v$ is a $k$-dimensional vector where $k$ is the number of
% features. The $i$th element is $1$ if the feature $\feat_i$ matches
% to the program. 
Checking whether the slice for the query includes the feature is done in the following two
steps. 

% \begin{enumerate}
% \item 

We first represent the query in Figure \ref{fig:overview}(a) itself
        by an abstract data-flow graph:
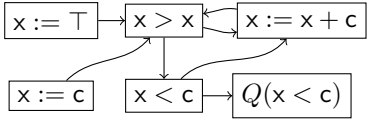
\begin{center}
\begin{tikzpicture}
  \node[draw] (Assign1) at (0,0) {${\sf x} := {\sf c}$};
  \node[draw] (Assume) at (1.5,0)
  {${\sf x} < {\sf c}$};
  \node[draw,fill=white,text=black] (Query) at (3.2,0) {$Q({\sf
      x} < {\sf c})$};
  \node[draw] (Assign2) at (3.33,1) {${\sf x} := {\sf x } + {\sf c}$};  
  \node[draw] (GT) at (1.5,1) {${\sf x} > {\sf x }$}; 
  \node[draw] (Assign3) at (0,1) {${\sf x} := \top$};  
  \draw[->] (Assign1) to[in=230] (GT);
  \draw[->] (Assume) to (Query);
  \draw[->] (Assume) to[in=230] (Assign2);
  \draw[->] (GT) to (Assume);
  \draw[->] (Assign2) to[in=10,out=170] (GT);
  \draw[->] (GT) to[in=190,out=-10] (Assign2);
  \draw[->] (Assign3) to (GT);
\end{tikzpicture}
\vspace*{-1.5mm}
\end{center}
Note that the graph is similar to the one in Figure~\ref{fig:overview}(c) 
but it contains all the parts of the original 
program. For instance, it has the node ${\sf x}>{\sf x}$
and the edge from this node to ${\sf x}<{\sf c}$, both of which are
absent in the feature.  The unknown value, such as the return
value of {\tt unknown()}, is represented by $\top$. 

Next, we use a variant of graph inclusion to decide whether the query includes the feature.  
        We check whether every vertice of the feature is included in the graph of 
        the query and whether every arc of the feature is included in the \emph{transitive 
        closure} of the graph. The answers to both questions are yes.
        For instance, the path for 
        the arc $\fbox{{\sf x}:={\sf x}+{\sf c}} \to \fbox{{\sf
            x}$<${\sf c}}$ in the feature
        is $\fbox{{\sf x}:={\sf x}+{\sf c}} \to \fbox{{\sf x}$>${\sf x}} \to \fbox{{\sf
    x}$<${\sf c}}$ in the graph of the query.
%\end{enumerate}

Note that we use a variant of graph inclusion where
an arc of one graph is allowed to be realized by a \emph{path} of its including
graph, not necessarily by an \emph{arc} as in the usual definition. 
This variation is essential for our purpose. When we check
a feature against a query, the feature is reduced but the query is not.
Thus, even when the query here is the one from which the feature is generated,
this checking is likely to fail if we use the usual notion of graph inclusion
(i.e.,  $G_1=(V_1,E_1)$ is included in
$G_2 = (V_2,E_2)$ iff $V_1 \subseteq V_2$ and $E_1 \subseteq E_2$).
In theory, we could invoke a reducer on the query, but
this is not a viable option because reducing is just too expensive to perform 
every time we analyze a program. Instead, we take a (less expensive) alternative
based on the transitive closure of the graph of the query.

\section{Setting}
\label{sec:staticanalysis} 

\paragraph{Parametric Static Analysis}
We use the setting for parametric static
analyses in \cite{LiangTN11}.  Let $P \in \mbp$ be a program to analyze.
We assume that a set $\query_P$ of queries (i.e., assertions) in $P$
is given together with the program.  The goal of the analysis is to
prove as many queries as possible.  A static analysis is parameterized
by a set of program components. We assume a set $\component_P$ of
program components that represent parts of $P$. For instance, in our
partially flow-sensitive analysis, $\component_P$ is the set
of program variables.  The parameter space is defined by
$(\param_P, \sqsubseteq)$ where $\param_P$ is the binary vector
$\vec{a} \in \param_P = \mbb^{\component_P} = \myset{0,1}^{\component_P}$ with the pointwise
ordering.
% $\vec{a} \sqsubseteq \vec{a'} \iff \forall j \in \component_P.\;
% \vec{a}_j \leq \vec{a'}_j$.
We sometimes regard a parameter $\vec{a} \in \mca_P$ as a function
from $\component_P$ to $\mbb$, or the set
$\vec{a} = \myset{\idx \in \component_P \mid \vec{a}_{\idx} = 1}$.  In
the latter case, we write $|\vec{a}|$ for the size of the set.  We
define two constants in $\mca_P$:
$\vec{0} = \lambda j \in \component_P.\, 0$ and
$\vec{1} = \lambda j \in \component_P.\, 1$,
which represent the most
imprecise and precise abstractions, respectively.  
We omit the subscript $P$ when there is no confusion.
A parametric static analysis is a function
$\aF: \Pgm \times \param \to \power{\query}$, which
takes a program to analyze and a parameter, and returns a set of
queries proved by the analysis under the given parameter.
 \longversion{In static analysis of C
programs, using a more refined parameter typically improves the
precision of the analysis but increases the cost.
}

\vspace*{-0.5mm}
\paragraph{Analysis Heuristic that Selects a Parameter}
The parameter of the analysis is selected
by an analysis heuristic $\strategy: \Pgm \to \mca$.
Given a program $P$, the analysis first applies
the heuristic to $P$, and then uses the resulting parameter
$\strategy(P)$ to analyze the program. That is,
it computes $\aF(P, \strategy(P))$. 
If the heuristic is good, running the analysis with
$\strategy(P)$ would give results close to those of the most precise
abstraction ($\aF(P,\vec{1})$), while the analysis cost is close
to that of the least precise abstraction
($\aF(P,\vec{0})$). Previously, such a heuristic was
designed manually (e.g.,~\cite{Kastrinis2013,OhLHYY14,XinNY13}), which requires a
large amount of engineering efforts of analysis designers.

\section{Learning an Analysis Heuristic}
\label{sec:learning}

In a data-driven approach, an analysis heuristic $\strategy$ is
automatically learned from given codebases. In this section, we
describe our realization of this approach while assuming that a set of
features is given; this assumption will be discharged in
Section~\ref{sec:feat-generation}. We denote 
our method for learning a heuristic by $\Learn(F, \Feat, \vec{P})$,
which takes a static analysis $F$, a set $\Feat$ of features, and 
codebases $\vec{P}$, and returns a heuristic~$\strategy$.

\paragraph{Learning a Classifier}

In our method, learning a heuristic $\strategy$ boils down to
learning a classifier $\classifier$, which predicts for each query
whether the query can be proved by a static analysis with increased
precision but not without it.
% Thus, we aim to learn a classifier $\classifier$ from a
% given codebase. 
Suppose that we are given codebases $\vec{P} =
\myset{P_1,\dots,P_n}$, a set of features $\Feat =
\myset{\feat_1,\dots,\feat_k}$, and a procedure $\fpred$.
The precise definitions of $\Feat$ and $\fpred$
will be given in the next section. For now, it is sufficient
just to know that a feature $\feat_i \in \Feat$ % is a directed graph labeled
% with commands, and
describes a property about queries % that can be proved by an analysis 
% but only under high precision. Also,
% it suffices to know that
and $\fpred$ % uses a 
% variant of graph inclusion and
checks whether a query 
satisfies this property.

Using $\Feat$ and $\fpred$, we represent a query $q \in \query_P$ 
in a program $P$ by a feature vector $\Feat(q,P) \in \mbb^k$ such that
the $i$th component of the vector is the result of $\fpred(\feat_i,q,P)$.
% \[
% \Feat(q, P) = \langle \fpred (\feat_1,q,P), \dots, \fpred(\feat_k,q,P) \rangle.
% \]
This vector representation enables us to employ the standard tools for learning
and using a binary classifier. In our case, a classifier is just a map 
$\classifier: \mbb^k \to \mbb$ and predicts whether the query can be 
proved by the analysis under high precision (such as flow-sensitivity) 
but not with low precision (such as flow-insensitivity). To use such a classifier, 
we just need to call it with $\Feat(q,P)$. To learn it from codebases,
we follow the two steps described below:
\begin{enumerate}
        \item We generate labeled data $D \,{\subseteq}\, \mbb^k {\times} \mbb$ from the codebases:
          $D = \myset{\langle \Feat(q,P_i), b_i\rangle \mid P_i \in \vec{P} 
          \myland q \in \query_{P_i}}$, where $b_i = (q \in F(P_i,\vec{1}) \setminus F(P_i,\vec{0}))$. 
  That is, for each program $P_i \in \vec{P}$ and a query $q$ in
  $P_i$, we represent the query by a feature vector and label it with
  $1$ if $q$ can be proved by the analysis under the most precise setting
  but not under the least precise setting. When it is infeasible to run the most precise analysis
   (e.g., the Octagon analysis), we instead run an approximate version of it.
   In experiments with the partial Octagon analysis, we used the impact
   pre-analysis~\cite{OhLHYY14} as an approximation.

  % \hy{Can we really generate such training data? Isn't it too expensive
  % to run the analysis with the highest precision to all queries
  % in all programs in the training set?}

\item Then, we learn a classifier from the labeled data $D$
        by invoking an off-the-shelf learning algorithm,
        such as logistic regression, decision tree, and support vector machine.
\end{enumerate}

\paragraph{Building an Analysis Heuristic}
We construct an analysis heuristic 
$\strategy: \mbp \to \mca$ 
from a learned classifier $\classifier$ as follows:
\vspace*{-1.5mm}
\[
\strategy (P) = \bigcup \myset{\dep(q) \mid q \in \query_P \myland
  \classifier(\Feat(q, P)) = 1 }
\vspace*{-1.5mm}
\]
The heuristic iterates over every query $q \in \query_P$ in the
program $P$, and selects the ones that get mapped to $1$
by the classifier $\classifier$. For each of these selected queries,
the heuristic collects the parts of $P$ that may affect
the analysis result of the query. This collection is done
by the function $\dep: \query \to \mca$, which 
satisfies that $q \in F(P,\vec{1}) \implies q \in F(P,\dep(q))$ for all queries
$q$ in $P$.
% Hakjoo: I think it is conceptually okay to say this
% \kwonsoo{Can we say $req$ always satisfies the condition? Yes
% in theory, but it is possible that the designed (by analysis designers) $req$ fails to do so, because of the limitation of long dependency chains} 
This function should be specified by an analysis designer,
but according to our experience, this is rather a straightforward task.
For instance, our instance analyses (namely, two partially flow-sensitive analyses 
and partial Octagon analysis) implement $\dep$ via a simple dependency analysis. For instance,
in our partially flow-sensitive analysis, $\dep(q)$ is
just the set of all program variables in the
dependency slice of $P$ for the query $q$. The result of $\strategy(P)$ is
the union of all the selected parts of $P$.

\section{Automatic Feature Generation}
\label{sec:feat-generation}

%The effectiveness of the learning algorithm in the previous section
%crucially depends on the choice of the features.  In this
%section, we present our framework for automatically generating
%features relevant to a given analysis task.

We now present the main contribution of this paper, our feature generation algorithm.
The algorithm first generates so called
feature programs from given codebases (Section~\ref{sec:generate-features}),
and then converts all the generated programs to abstract data-flow graphs
(Section~\ref{sec:representing-features}). The obtained graphs enable
the $\fpred$ procedure 
to transform queries to feature vectors so that a classifier can be applied
to these queries (Section~\ref{sec:matching-features}). 
\longversion{In this section,
we will explain all these aspects of our feature generation algorithm.}
% In our
% approach, a feature corresponds to an abstract data-flow graph whose
% abstraction level is determined by a subset $\hat{R}$ of grammar rules
% for the target programming language.  
%In Section
%\ref{sec:final-algorithm}, we present the final feature generation
%algorithm. 
% that also automates the process of finding a right
% abstraction parameter $\hat{R}$ via cross-validation. 

\subsection{Generation of Feature Programs}
\label{sec:generate-features}

% assume that each program $P \in \vec{P}$ in the codebase contains a
% single query. Let $q_P$ be the query in program $P$. 

Given a static analysis $F$ 
and codebases $\vec{P}$,
% \footnote{In our experiments,
% we partitioned the codebases into two groups, the first
% being used for generating features and learning heuristics
% and the other for testing the generated features 
% and the learned heuristics.}
$\GenerateFP(\vec{P}, F)$ generates feature programs in two steps.

First, it collects the set of 
queries in $\vec{P}$ that can be proved
by the analysis $F$ under high precision (i.e., $\StaticAnalyzer(-,\vec{1})$) but not with low precision
(i.e., $\StaticAnalyzer(-,\vec{0})$).  
We call such queries \emph{positive} and the other non-positive queries \emph{negative}. 
The negative
queries are either too hard in the sense that they cannot be proved even with high precision,
or too easy in the sense that they can be proved even with low precision.
Let $\positive$ be the set of positive queries and their host
programs:
\[
\openup-1.4\jot
\begin{aligned}
\positive = \myset{(P,q) \mid P \in \vec{P} \myland q \in \mbq_P
  \myland \reducecond_q (P) } 
\end{aligned}
\vspace*{-1.4mm}
\]
where $\mbq_P$ is the set of queries in $P$ and 
$\reducecond_q$ is defined by:
% \footnote{When 
% it is infeasible to run $\StaticAnalyzer(P,\vec{1})$ as in the case of partial Octagon, 
% we use instead its approximation such as the impact pre-analysis in \cite{OhLHYY14}.}
\begin{equation}
\openup-1.3\jot
\label{eq:cond}
  \reducecond_q(P) = \left(q \not\in \StaticAnalyzer(P,\vec{0})  \myland q
  \in \StaticAnalyzer(P,\vec{1})\right).
\vspace*{-1.3mm}
\end{equation}
%The predicate $\reducecond_q(P)$ holds when the analysis $F$ with
%low precision cannot prove the query $q$ (i.e., $q \not\in
%\StaticAnalyzer(P,\vec{0})$) but $F$ with high precision
%can (i.e., $q \in \StaticAnalyzer(P,\vec{1})$).

Second, $\GenerateFP(\vec{P}, F)$ shrinks the positive queries 
collected in the first step by using a program reducer.
A program reducer (e.g., C-Reduce~\cite{creduce}) is a function
of the type: $\reducer: \Pgm \times (\Pgm \to \mbb) \to \Pgm$.
It takes a program $P$ and a predicate $\pred$, and removes parts of $P$
as much as possible while preserving the original result of the
predicate. At the end, it returns a minimal program $P'$ such that 
$\pred(P') = \pred(P)$. Our procedure $\GenerateFP(\vec{P}, F)$ runs
a reducer and shrinks programs in $\positive$ as follows:
\vspace*{-1.5mm}
\[
\positiveminimal = \myset{(\reducer(P,\reducecond_q), q) \mid (P,q)
                     \in \positive}.
\vspace*{-1.5mm}
\]
$\positiveminimal$ is the collection of
the reduced programs paired with queries. We call these programs
\emph{feature programs}. Because of the reducer, each feature program
contains only those parts related to the reason that high precision
is effective for proving its query. % , or that high precision is not useful
% for proving the query. 
Intuitively, the reducer removes noise in the
positive examples $(P,q) \in \positive$, % and negative examples
% $(P,q) \in \negative$,
until the examples contain only the reasons that 
high precision of the analysis helps prove their queries. % and why
% it does not help other queries.
The result of $\GenerateFP(\vec{P}, F)$ is $\positiveminimal$. % the union of $\positiveminimal$
% and $\negativeminimal$.

% \hy{What is written here is not quite consistent with what we say in the overview section.
% There we said that we are reducing only positive examples.}

\paragraph{Improvement 1: Preserving Analysis Results}
\label{sec:preserving-features}
A program reducer such as C-Reduce~\cite{creduce} is
powerful and is able to reduce C programs of thousands LOC
to just a few lines of feature programs. However, some additional
care is needed in order to prevent C-Reduce from removing too aggressively
and producing trivial programs. 

For example, suppose we
analyze the following code snippet (excerpted and simplified from {\tt bc-1.06}) with 
a partially flow-sensitive interval analysis:
\begin{Verbatim}[numbers=left]
yychar = 1; yychar = input(); //external input
if (yychar < 0) exit(1);
if (yychar <= 289)
  assert(0 <= yychar < 290); // query q
yychar++;
\end{Verbatim}
The predicate $\reducecond_q$ in \refeq{eq:cond} holds for this program.
The analysis can prove the assertion at line 4 with flow-sensitivity,
because in that case, it computes the interval $[0,289]$ for {\tt yychar} at line 4.
But it cannot prove the assertion with flow-insensitivity, because it computes
the interval $[-\infty,+\infty]$ for {\tt yychar} that holds over the entire program.

Reducing the program under the predicate $\reducecond_q$ may produce the following program:
\begin{verbatim}
yychar=1; assert(0<=yychar<290); yychar++;
\end{verbatim}
It is proved by flow-sensitivity but not by
flow-insensitivity. An ordinary flow-insensitive interval analysis
%uses widening and 
computes the interval $[1,+\infty]$ because of the
increment of {\tt yychar} at the end. 
Thus the resulting program still satisfies 
$\reducecond_q$. However, this reduced program does not contain 
the genuine reason that the original program needed flow-sensitivity:
in the original program, the \texttt{if} commands at lines 2 and 3 are analyzed 
accurately only under flow-sensitivity, and the accurate analysis of these
commands is crucial for proving the assertion.

To mitigate the problem, we run the reducer with a stronger predicate
that additionally requires the preservation of analysis result. In the flow-sensitive
analysis of our original
program, the variable {\tt yychar} has the interval value $[0,289]$ at
the assertion. In the above reduced program, on the other hand, it 
has the value $[1,1]$.  The strengthened predicate $\reducecond'_q$ in
this example is:
\vspace*{-1mm}
\begin{equation}
\label{eq:new-cond}
\reducecond'_q (P) = (\reducecond_q(P) \myland \mbox{value of {\tt yychar} at $q$ is $[0,289]$}).
\vspace*{-1mm}
\end{equation}
Running the reducer with this new predicate results in:
\vspace*{-1mm}
\begin{Verbatim}[numbers=left]
yychar = input();
if (yychar < 0) exit(1);
if (yychar <= 289) assert(0 <= yychar < 290);
\end{Verbatim}
The irrelevant commands
({\tt yychar = 1}, {\tt yychar++}) in the original program are removed by the reducer, but 
the important \texttt{if} commands at lines 2 and 3 remain in the reduced
program. Without these \texttt{if} commands, it is impossible to satisfy
the new condition
\refeq{eq:new-cond}, so that the reducer has to preserve them
in the final outcome. % For a negative query $q$ in a program $P$, 
% we similarly use the condition that $\neg \reducecond_q(P)$ and 
% the value of a variable involved in the query is preserved.
This idea of preserving the analysis result
during reduction was essential to generate diverse feature programs.
Also, it can be applied to any program analysis easily. 

% This idea of using a stronger condition played a key role to generate
% diverse feature programs in our flow-sensitive and relational octagon
% analyses. 

% \hy{The idea of preserving the analysis result during reduction is general,
% so it can be applied to any program analysis. But why should one expect
% that this idea makes things better? Can we come up with some general
% intuitive explanation?}

\paragraph{Improvement 2: Approximating Variable Initialization}

%\iffalse
%\kihong{Can we use another term rather than ``abstracting''? 
%We also use the term ``abstracting'' in abstract data flow graph.}
Another way for guiding a reducer is to replace commands in
a program by their overapproximations and to call the reducer
on the approximated program. The rationale is that approximating
commands would prevent the reducer from accidentally identifying
a reason that is too specific to a given query and does not generalize well.
%an analysis with high precision performs well
%for a given query. 
Approximation would help remove such reasons,
so that the reducer is more likely to find general reasons.
%\kwonsoo{The two sentences above are contradicting each other.}
%\kihong{The above paragraph is unclear and misleading for me. 
%I don't think ``the reducer identifies a wrong reason'' or 
%``the reducer makes a mistake''. 
%The reducer just finds a weak (or not that useful) reason.}

Consider the following code snippet (from {\tt spell-1.0}):
\begin{Verbatim}[numbers=left]
pos = 0;
while (1) { if (!pos) assert(pos==0); pos++; }
\end{Verbatim}
The flow-sensitive interval analysis proves the assertion because
it infers the interval $[0,0]$ for {\tt pos}.
Note that this analysis result crucially relies on the condition {\tt !pos}.
Before the condition, the value of {\tt pos} is $[0,+\infty]$,
but the condition refines the value to $[0,0]$. However, reducing the program under
$\reducecond_q$ in \eqref{eq:cond} leads to the following program:
\begin{verbatim}
pos=0; assert(pos==0); pos++;
\end{verbatim}
%\kihong{I don't understand why the reducer with condition (1) results in the above program.
%The query can be proven by even a flow-insensitive analysis}
This reduced program no longer says the importance
of refining an abstract value with the condition {\tt !pos}.
Demanding the preservation of the analysis result does not help,
because the value of {\tt pos} is also $[0,0]$ in the reduced
program. 

We fight against this undesired behavior of the reducer
by approximating commands of a program $P$ before passing it to the reducer. 
Specifically, for every positive query $(P,q)$ and for each command in $P$ 
that initializes a variable with a constant value, 
we replace the constant by Top (an expression that denotes 
the largest abstract value $\top$) as long as this replacement
does not make the query negative. For instance, we transform our example 
to the following program:
\begin{verbatim}
pos = Top;  // 0 is replaced by Top
while (1) { if (!pos) assert(pos==0); pos++; }
\end{verbatim}
Note that {\tt pos = 0} is replaced by {\tt
pos = Top}. Then, we apply the reducer to this transformed program, 
and obtain:
\begin{verbatim}
pos = Top; if (!pos) assert(pos==0);
\end{verbatim}
Note that the reduced program keeps the condition {\tt !pos};
because of the change in the initialization of {\tt pos},
the analysis cannot prove the assertion without using the condition {\tt !pos} 
in the original program. 

\subsection{Transformation to Abstract Data-Flow Graphs}
\label{sec:representing-features}

Our next procedure is $\GenerateDFG(\positiveminimal, \hat{R})$, which
converts feature programs in $\positiveminimal$ to their data-flow
graphs where nodes are labeled with the abstraction of
atomic commands in those programs. We call such graphs 
\emph{abstract data-flow graphs}. These graphs act
as what people call features in the applications of machine learning.
The additional parameter
$\hat{R}$ to the procedure controls the degree of abstraction
of the atomic commands in these graphs. A method for finding an appropriate
parameter $\hat{R}$ will be presented in Section~\ref{sec:final-algorithm}.

%Below, we define the
%abstract commands and how to transform feature programs into
%data-flow graphs. Finding the right parameter $\hat{R}$ automatically will be
%explained in Section \ref{sec:final-algorithm}. 

\paragraph{Step 1: Building Data-Flow Graphs}
The first step of $\GenerateDFG(\positiveminimal, \hat{R})$ is to build
data-flow graphs for feature programs in $\positiveminimal$ and to
slice these graphs with respect to queries in those programs.

The $\GenerateDFG$ procedure constructs and slices such data-flow graphs
using standard recipes. Assume a feature program $P \in \positiveminimal$ represented
by a control-flow graph $(\mbc,\cfgto)$, where
$\mbc$ is the set of program points annotated with atomic commands
and $(\cfgto) \subseteq \mbc \times \mbc$ the control-flow
relation between those program points. The data-flow graph for $P$ reuses the node set $\mbc$ of the
control-flow graph, but it uses a new arc relation $\dugto$:
$c \dugto c'$ iff there is a def-use chain in $P$ from $c$ to $c'$ 
on a memory location or variable $l$ (that is, $c \cfgto^+  c'$, $l$ is defined at $c$, $l$ is used 
at $c'$, and $l$ is not re-defined in the intermediate program points between $c$ and $c'$). 
For each query $q$ in the program $P$, its slice $(\mbc_q, \dugto_q)$ is just
the restriction of the data-flow graph $(\mbc_q, \dugto_q)$ with respect
to the nodes that may reach the query 
(i.e., $\mbc_q = \myset{c \in \mbc \mid c \dugto^* c_q}$).

\paragraph{Step 2: Abstracting Atomic Commands}

The second step of $\GenerateDFG(\positiveminimal, \hat{R})$
is to abstract atomic commands in the data-flow graphs obtained in the first step
and to collapse nodes labeled with the same abstract command.
This abstraction is directed by the parameter $\hat{R}$,
and forms the most interesting part of the 
$\GenerateDFG$ procedure.

\begin{figure}[t]
\vspace*{-4mm}
  \[
  \begin{array}{@{\!\!}lr@{~}l@{~}}
    R_1:& c&\to lv := e \mid lv := \alloccmd(e) \mid \assumecmd(e_1
             \compare e_2)  \\
    R_2:&  e  &\to {\sf c} \mid e_1 \binop e_2 \mid lv \mid \& lv,\quad lv \to {\sf x} \mid *e \mid e_1[e_2]  \\
    R_3:&  \binop &\to + \mid - \mid * \mid / \mid << \mid >>, \;\;
                    \compare\;\to\; < \mid \le \mid > \mid \ge \mid = \mid
                    \not= \\
    R_4:& {\sf c} & \to  0 \mid 1 \mid 2 \mid \cdots,~\quad\quad\qquad {\sf x}  \to x \mid y \mid z \mid \cdots 
  \end{array}
\vspace*{-3mm}
  \]
  \caption{The set $R$ of grammar rules for C-like languages% for expressions and atomic
                                % commands
}
  \label{fig:language}

\end{figure}

Our abstraction works on the grammar for the atomic commands
shown in Figure~\ref{fig:language}.  The command $lv:=e$ assigns the
value of $e$ into the location of $lv$, and $lv := \alloccmd(e)$
allocates an array of size $e$. The assume command
$\assumecmd(e_1\compare e_2)$ allows the program to continue only when the
condition evaluates to true. An expression may be a constant integer
(${\sf c}$), a binary operator ($e_1\binop e_2$), an l-value
expression ($lv$), or an address-of expression ($\& lv$). An l-value
may be a variable (${\sf x}$), a pointer dereference ($*e$), or an
array access ($e_1[e_2]$).
%We assume that the number of rules in the grammar is finite,
%which is possible by defining ${\sf c}$ and ${\sf x}$ to be the set of integer
%constants and variables that appear in some program in $\positiveminimal$.
%
%\hy{Double-check that we are not contradicting ourselves when we make the finiteness
%assumption above.}

Let $R$ be the set of grammar rules in Figure~\ref{fig:language}.
The second parameter $\hat{R}$ of $\GenerateDFG(\positiveminimal, \hat{R})$ is a subset
of $R$. It specifies how each atomic command should be abstracted. Intuitively,
each rule in $\hat{R}$ says that if a part of an atomic command matches the RHS of the rule,
it should be represented abstractly by the nonterminal symbol in the LHS of the rule. 
For example, when $\hat{R} = \myset{\binop \to +\,|\, -}$,  both $x=y+1$
and $x=y-1$ are represented abstractly by the same $x=y\binop 1$, where $+$ and $-$ are 
replaced by $\binop$. 
Formally, $\GenerateDFG(\positiveminimal, \hat{R})$ transforms the parse tree of each atomic
command in $\positiveminimal$ by repeatedly applying the grammar rules in $\hat{R}$ backwards
to the tree until a fixed point is reached. This transformation loses
information about the original atomic command, such as the name of a binary operator. We
denote it by a function $\alpha_{\hat{R}}$. The following example illustrates this transformation
using a simplified version of our grammar.
\begin{example}
Consider the grammar:
$R = \myset{
e \to {\sf x} \mid {\sf c} \mid e_1 \binop e_2,\quad {\sf x} \to x \mid y,
\quad {\sf c} \to 1 \mid 2,\quad \binop \to + \mid -}$.
Let $\hat{R} = \myset{{\sf x} \to x\;|\; y,~ {\sf c} \to 1\;|\;2,~ \binop \to + \;|\; - }$. Intuitively, $\hat{R}$
specifies that we should abstract variables, constants, and operators in atomic commands
and expressions by nonterminals ${\sf x}$, ${\sf c}$, and $\binop$, respectively.  The abstraction is done by applying
rules $\hat{R}$ backwards to parse trees until none of the rules becomes applicable. 
For example, the expression
$x+1$ is abstracted into ${\sf x}\binop {\sf c}$  as follows:
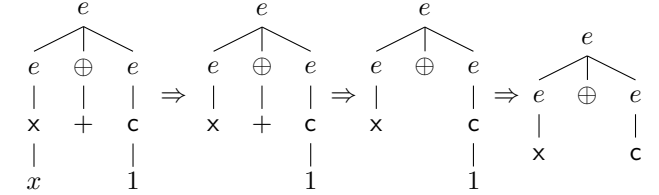
\vspace*{-2.7mm}
 \[
\begin{array}{c}
\Tree [.$e$ [.$e$ [.${\sf x}$ $x$ ] ] [.$\binop$ $+$ ] [.$e$ [.${\sf
  c}$ $1$ ] ] ]
\end{array}
\Rightarrow
\begin{array}{c}
\Tree [.$e$ [.$e$ ${\sf x}$ ] [.$\binop$ $+$ ] [.$e$ [.${\sf c}$ $1$ ]
  ] ]
\end{array}
{\Rightarrow}
\begin{array}{c}
\Tree [.$e$ [.$e$ ${\sf x}$ ] $\binop$ [.$e$ [.${\sf c}$ $1$ ]
  ] ]
\end{array}
{\Rightarrow}
\begin{array}{c}
\Tree [.$e$ [.$e$ ${\sf x}$ ] $\binop$ [.$e$ ${\sf c}$
  ] ]
\end{array}
\vspace*{-2.6mm}
\]
We first apply the rule ${\sf x} \to x$ backwards to the parse tree
(leftmost) and collapse the leaf node $x$ with its parent. Next, we
apply $\binop \to +$ where $+$ is collapsed to $\binop$. Finally, we
apply the rule ${\sf c} \to 1$, getting the rightmost tree. The
result is read off from the last tree, and is the abstract 
expression ${\sf x} \binop {\sf c}$. Similarly, 
$y-2$ gets abstracted to ${\sf x} \binop {\sf c}$.
%(according to $\hat{R}$). 
$\qed$
\end{example}

For each data-flow slice computed in the first step,
our $\GenerateDFG(\positiveminimal, \hat{R})$ procedure applies
the $\alpha_{\hat{R}}$ function to the atomic commands
in the slice. Then,
it merges nodes in the slice to a single node if
they have the same label (i.e., the same abstract atomic command).
The nodes after merging inherit the arcs from the original
slice. We call the resulting graphs 
%labeled with (abstract)
%atomic commands are called 
\emph{abstract data-flow graphs}.
%and correspond to so called features in the application
%of machine learning techniques. 
These graphs describe (syntactic and semantic) 
program properties, such as the ones used in~\cite{OhYaYi15}. For example, 
the abstract data-flow graph
$({\sf x} < {\sf c}) \dugto ({\sf x} := \alloccmd ({\sf x}))$
says that a program variable is
compared with a constant expression before being used as an argument of a
memory allocator (which corresponds to the features \#9 and \#11 for
selective flow-sensitivity in~\cite{OhYaYi15}).

We write $\{\pi_1,\ldots,\pi_k\}$ for the abstract data-flow
graphs generated by $\GenerateDFG(\positiveminimal, \hat{R})$. We sometimes
call $\pi_i$ \emph{feature}, especially when we want to emphasize its role in
our application of machine learning techniques.

%\paragraph{Finding the Right Abstraction Level}
%Before moving on, 
We point out that the performance of a data-driven analysis in our
approach depends on the
choice of the parameter $\hat{R}$ to the $\GenerateDFG(\positiveminimal, \hat{R})$ procedure.
For example, the Octagon analysis can track certain binary operators such as addition precisely
but not other binary operators such as multiplication and shift. 
%it has difficulties in inferring accurate information about other binary operators such as multiplication
%and shift. 
Thus, in this case, we need to use $\hat{R}$
that at least differentiates these two kinds of operators. 
In Section~\ref{sec:final-algorithm}, we describe 
a method for automatically choosing $\hat{R}$ from data via iterative cross validation.% ,
% and describe how it fits into the final feature-generation algorithm.

\subsection{Abstract Data-flow Graphs and Queries}
\label{sec:matching-features}

Abstract data-flow graphs encode properties about queries.
These properties are checked by our $\fpred_{\hat{R}}$ procedure
parameterized by $\hat{R}$. The procedure
takes an abstract data-flow graph $\feat$, a query $q$
and a program $P$ that contains the query. Given such inputs, 
it works in four steps. First, $\fpred_{\hat{R}}(\feat,q,P)$ normalizes
$P$ syntactically so that some syntactically different yet
semantically same programs become identical. Specifically, the procdure
eliminates temporary variables (e.g., convert {\tt tmp = b + 1; a = tmp;} to {\tt a = b +
  1}), removes double negations (e.g., convert {\tt assume
  (!(!(x==1)))} to {\tt assume (x==1)}), and makes explicit conditional
expressions (e.g., convert {\tt assume(x)} to {\tt assume(x!=0)}).
Second, $\fpred_{\hat{R}}(\feat,q,P)$ constructs
a data-flow graph of $P$, and computes the slice of the graph that may
reach $q$. Third, it builds an abstract data-flow graph from this slice.
That is, it abstracts the atomic commands in the slice,
merges nodes in the slice that are labeled with the
same (abstract) atomic command, and induces arcs between nodes after
merging in the standard way. Let $(N_q, \dugto_q)$ be the resulting abstract
data-flow graph, and $(N_0, \dugto_0)$ the node and arc sets 
of $\feat$. In both cases, nodes are identified with their labels,
so that $N_q$ and $N_0$ are the sets of (abstract) atomic commands.
Finally, $\fpred_{\hat{R}}(\feat,q,P)$ returns $0$ or $1$ according to the 
criterion: $\fpred_{\hat{R}}(\feat,q,P) = 1
        \iff  
        N_0 \subseteq N_q \land (\dugto_0) \subseteq (\dugto_q^*)$. 
\longversion{The criterion means that the checking of our procedure succeeds if all the 
atomic commands in $N_0$ appear in $N_q$ and their dependencies
encoded in $\dugto_0$ are respected by the transitive dependencies
$\dugto_q^*$ in the query.}
 Taking the transitive closure $(\dugto_q^*)$ here 
is important. It
enables $\fpred_{\hat{R}}(\feat,q,P)$ to detect whether the programming pattern 
encoded in $\feat$ appears somewhere in the program slice for $q$,
even when the slice contains commands not related
to the pattern.

\begin{algorithm}[t]
\caption{Automatic Feature Generation}
\label{alg:overall}
\begin{algorithmic}[1]
\Require codebases $\vec{P}$, static analysis
$F$, grammar rules $R$
\Ensure a set of features $\Feat$ 

\State partition $\vec{P}$ into $\vec{P}_{\it tr}$ and $\vec{P}_{\it va}$
\Comment{training/validation sets}
\State $\positiveminimal \gets {\GenerateFP} (\vec{P}_{\it tr}, F)$
\Comment{generate feature programs}
%\State partition $\vec{P}$ into $\vec{P}_{\it tr}$ and
%$\vec{P}_{\it va}$ \Comment{training/validation sets}
\State $s_{\it best}, \Feat_{\it best} \gets -1, \emptyset$
\Repeat
  \State $\hat{R} \gets$ choose a subset of $R$ (i.e., $\hat{R}
  \subseteq R$)
  \State $\Feat \gets \GenerateDFG(\positiveminimal, \hat{R})$
  \Comment{Build data-flow graphs}
  \State $\strategy_{\classifier} \gets \Learn (F, \Feat, \vec{P}_{\it
    tr})$
  \State $s \gets \evaluate(F, {\classifier}, \vec{P}_{\it va})$
  \Comment{Evaluate $F_1$-score of $\classifier$}
   \If{$s > s_{\it best}$}
   \State $s_{\it best}, \Feat_{\it best} \gets s, \Feat$
   \EndIf

\Until~timeout
\State \textbf{return} $\Feat_{\it best}$
\end{algorithmic}
\end{algorithm}

\subsection{Final Algorithm}
\label{sec:final-algorithm}

Algorithm~\ref{alg:overall} shows the final algorithm for feature
generation.
%, which puts together all the processes explained in this
%section.  
It takes codebases
$\vec{P} = \myset{P_1,\dots,P_n}$, a static analysis
$\StaticAnalyzer$ (Section \ref{sec:staticanalysis}), and a set $R$ of
grammar rules for the target programming language. Then, it returns
the set $\Feat$ of features. % (i.e., abstract
% data-flow graphs).  that are generated for the given static analysis
% and the verification task.

% \hy{We say that $r/n$ is around $0.6$. The word ``around'' looks
% strange to reviewers. It appeared because of partial Octagon.  Why
% didn't we just use the same number?}

The algorithm begins by splitting the codebases $\vec{P}$ into a
training set $\vec{P}_{\it tr} = \myset{P_1,\dots,P_r}$ and a
validation set $\vec{P}_{\it va} = \myset{P_{r+1},\dots,P_n}$ (line
1).  In our experiments, we set $r$ to the nearest integer to
$0.7n$. Then, the algorithm calls $\GenerateFP$ with
$\vec{P}_{\it tr}$ and the static analysis, so as to generate feature
programs.  Next, it initializes
the score $s_{\it best}$ to $-1$, and the set of features
$\Feat_{\it best}$ to the empty set. At lines 4--12, the algorithm
repeatedly improves $\Feat_{\it best}$ until it hits the limit of the
given time budget. Recall that the performance of our approach
depends on a set $\hat{R}$ of grammar rules, which determines
how much atomic commands get abstracted. 
 In each iteration of the
loop, the algorithm chooses $\hat{R} \subseteq R$ according to the
strategy that we will explain shortly, and calls
$\GenerateDFG(\vec{P}_{\it feat}, \hat{R})$ to generate a new
candidate set of features $\Feat$. Then, using this candidate set, the
algorithm invokes an off-the-shelf learning algorithm (line 7) for learning an
analysis heuristic $\strategy_{\classifier}$ from the training data
$\vec{P}_{\it tr}$; the subscript $\classifier$ denotes a
classifier built by the learning algorithm.  The quality of the
learned heuristic is evaluated on the validation set
$\vec{P}_{\it va}$ (line 8) by computing the
$F_1$-score\footnote{$2\cdot {\mbox{precision}\cdot
    \mbox{recall}}/({\mbox{precision}+\mbox{recall}})$.}
of $\classifier$. 
If this evaluation gives a better score than
the current best $s_{\it best}$, the set $\Feat$ becomes a new current
best $\Feat_{\it best}$ (lines 9--10).  
To save computation, before running our algorithm,
we run the static analysis $F$ for all programs in the codebases $\vec{P}$
with highest precision $\vec{1}$ and again with lowest precision $\vec{0}$,
and record the results as labels for all queries in $\vec{P}$. 
This preprocessing lets us avoid calling $F$ in $\Learn$ and $\evaluate$.
\longversion{As a result, after feature programs
$\positiveminimal$ are computed at line 2, building data-flow
graphs and learning/evaluating the heuristic do not invoke the
static analysis, so that each iteration of the loop in Algorithm
\ref{alg:overall} runs fast. }

%\hy{Add the citation on the F-measure. I don't know what it is.}

Our algorithm chooses a subset $\hat{R} \subseteq R$ of grammar
rules using a greedy bottom-up search. It partitions the
grammar rules in Figure \ref{fig:language} into four groups 
$R = R_1 \uplus R_2 \uplus R_3 \uplus R_4$ such that $R_1$ contains
the rules for the nonterminal $c$ for commands,
$R_2$ those for the nonterminals $e,lv$ for expressions,
$R_3$ the rules for the nonterminals $\binop,\compare$ for operators,
and $R_4$ those for the remaining nonterminals ${\sf x}, {\sf c}$
for variables and constants. These sets form a hierarchy with $R_i$ above $R_{i+1}$ for $i \in \myset{1,2,3}$ in the following sense: for a typical
derivation tree of the grammar, an instance
of a rule in $R_i$ usually appears nearer to the root of the tree than
that of a rule in $R_{i+1}$. 
Algorithm \ref{alg:overall} begins by choosing a subset of $R_3$ randomly
and setting the current rule set $\hat{R}$ to the union of this subset
and $R_4$. Including the rules in $R_4$ has the effect of making
the generated features (i.e., abstract data-flow graphs) forget
variable names and constants that are specific to programs
in the training set, so that they generalize well across different
programs. This random choice is repeated for a fixed number of times
(without choosing previously chosen abstractions),
and the best $\hat{R}_3$ in terms of its score $s$ is recorded.
Then, Algorithm~\ref{alg:overall} similarly tries different randomly-chosen
subsets of $R_2$ but this time using the best $\hat{R}_3$ found, instead
of $R_4$, as the set of default rules to include. The best choice
$\hat{R}_2$ is again recorded. Repeating this process with
$R_1$ and the best $\hat{R}_2$ gives the final result $\hat{R}_1$, which leads
to the result of Algorithm~\ref{alg:overall}. 
%Note that
%this strategy for choosing $\hat{R}$ of our algorithm is greedy
%in that once we find a best $\hat{R}_i$ at a level $i$, 
%we only add new grammar rules in the upper level $i-1$ to $\hat{R}_i$, 
%without deleting any from $\hat{R}_i$. 

% We start by building a set $\hat{R}$ using program constructs only at the
% lowest level within the hierarchy (e.g., $\hat{R}=\{+,*,<<\}$). When we exhaust the time budget for the level, we keep the best $\hat{R}$ and move on
% to the upper level. Our search is greedy in that once we have found a best
% $\hat{R}$ from a lower level, we only add new program constructs in the upper
% level into $\hat{R}$, without deleting any program constructs from lower
% levels. Each level is explored within a time budget of ?? and
% we set the size $k$ of $\hat{R}$ to be $1<=k<=6$.

% \kwonsoo{Our paper writes that we abstract away variable names.
% In implementation, however, we keep mappings between variable names from the feature-side and those from the target-query-side during the match process.}

%%% Local Variables:
%%% mode: latex
%%% TeX-master: "paper"
%%% End:

\section{Instance Analyses}
\label{sec:instances}
We have applied our feature-generation algorithm to three
parametric program analyses: partially flow-sensitive interval
and pointer analyses, and a partial Octagon analysis.
\longversion{
These analyses are designed following the data-driven approach of
Section~\ref{sec:learning}, so they are equipped with engines 
for learning analysis heuristics from given codebases. Our algorithm
generates features required by these learning engines.
}

In this section, we describe the instance analyses. 
%The results of our experiments with the analyses will be reported in Section~\ref{sec:experiments}.
In these analyses,
a program is given by its control-flow graph $(\mbc,\cfgto)$, where
each program point $c \in \mbc$ is associated with an atomic command in
Figure \ref{fig:language}.
We assume heap abstraction based on allocation sites
and the existence of a variable for each site in a program. This 
lets us treat dynamically allocated memory cells simply as variables.

\paragraph{Two Partially Flow-sensitive Analyses}

We use partially flow-sensitive interval and pointer analyses
that are designed according to the recipe in~\cite{OhYaYi15}.
These analyses perform the sparse analysis~\cite{OhHLLY12,Farzan12,sparse2014,controlfeasibility-oopsla15}
in the sense that they work on \emph{data-flow} graphs.
Their flow-sensitivity is controlled by a chosen set of program
variables; only the variables in the set are analyzed flow-sensitively.
In terms of the terminologies of Section \ref{sec:staticanalysis},
the set of program components $\component$ is 
that of variables $\Var$, an analysis
parameter $\vec{a} \in \mca = \myset{0,1}^\component$ specifies a subset
of $\Var$.% , and the most imprecise and precise parameters are
% $\vec{0} = \emptyset$ and $\vec{1} = \Var$.

Both interval and pointer analyses define functions $\StaticAnalyzer \colon 
\Pgm \times \mca \to \power{\Query}$ 
that take a program and a set of variables and return proved queries in the program. 
They compute mappings $D \in \mbd = \mbc \to \mbs$
from program points to abstract states, where
an abstract state $s \in \mbs$ itself is a map from program variables to values, i.e., $\mbs = \Var \to \mbv$.
In the interval analysis, $\mbv$ consists of intervals,
and in the pointer analysis, $\mbv$ consists of sets of the addresses of program variables.
% Defining the transfer function for
% the interval and pointer analyses for C is well-known and we omit the
% details. The analyses use the allocation-site-based abstraction for
% heap, array smashing, field-sensitivity, and
% context-insensitivity. Our pointer analysis is Andersen-style. 

Given an analysis parameter $\vec{a}$,
the analyses compute the mappings $D \in \mbd$ as follows. 
First, they construct a data-flow graph for variables in $\vec{a}$.
For each program point $c \in \mbc$,
let
$\Def(c) \subseteq {\Var}$ and $\Use(c) \subseteq {\Var}$ be the
definition and use sets. Using these sets, the analyses construct
a data-flow relation~$(\defusea_{\vec{a}}) \subseteq \mbc \times \Var \times
\mbc$: $c_0 \defuse{x}_{\vec{a}} c_n$ holds 
% \[
%         \begin{array}{@{}r@{\;}c@{\;}l@{}} 
%                 c_0 \defuse{x}_{\vec{a}} c_n 
%                 & \iff & \exists [c_0,\dots, c_n] \in \Paths.\,\exists x \in \vec{a}. 
%                 \\ 
%                 & & x \in \Def(c_0) \cap \Use(c_n) \land \forall 0 < i < n.\, x\not\in \Def(c_i).  
%         \end{array}
% \]
if there exists a path $[c_0, c_1,\dots, c_n]$ in the control-flow graph 
such that $x$ is defined at $c_0$ (i.e., $x \in \Def(c_0)$) and used
at $c_n$ (i.e., $x \in \Use(c_n)$), but it is not re-defined 
at any of the intermediate points $c_i$, and the variable $x$ is
included in the parameter $\vec{a}$. Second, the analyses perform flow-insensitive
analyses on the given program, and store the results in $s_I \in \mbs$.
Finally, they compute % exact or approximate
fixed points of the function $F_{\vec{a}}(D) = \lambda c.f_c(s')$
where $f_c$ is a transfer function 
at a program point $c$, and the abstract state $s'$ is the 
following combination of $D$ and $s_I$ at $c$: $s'(x) = s_I(x)$,
for $x \not\in \vec{a}$
and for $x \in \vec{a}$, $s'(x) = \bigsqcup_{c_0 \defuse{x}_{\vec{a}} c} D(c_0)(x)$.
% \[ s'(x) 
%         = \left\{\begin{array}{lr} 
%                 s_I(x) &  (x \not\in \vec{a}) 
%                 \\ 
%                 \bigsqcup_{c_0 \defuse{x}_{\vec{a}} c} D(c_0)(x) & \mbox{otherwise} 
%         \end{array}\right.
% \]
%using the standard iterative algorithm:
% \begin{align*} 
%         F_{\vec{a}}(D)  
%         & =  \lambda c.\; f_c(s') 
%         \\ 
%         & \mbox{where}\ s'(x) 
%         = \left\{\begin{array}{lr} 
%                 s_I(x) &  (x \not\in \vec{a}) 
%                 \\ 
%                 \bigsqcup_{c_0 \defuse{x}_{\vec{a}} c} D(c_0)(x) & \mbox{otherwise} 
%         \end{array}\right.
% \end{align*}
%  and 
% specifies how an interval or a pointer analysis handles the atomic
% command associated with $c$. 
Note that for variables not in
$\vec{a}$, $F_{\vec{a}}$ treats them flow-insensitively by using $s_I$.  When ${\vec{a}} = \Var$, 
the analyses become ordinary flow-sensitive analyses, and when
${\vec{a}} = \emptyset$, they are just flow-insensitive analyses.

\paragraph{Partial Octagon Analysis}

We use the partial Octagon analysis formulated
in~\cite{HeoOhYa16}.  Let $m$ be the number of variables in the program,
and write $\Var = \{x_1,\ldots,x_m\}$. The set of program components
$\component$ is $\Var \times \Var$, so an analysis parameter
$\vec{a} \in \mca = \myset{0,1}^\component$ consists of
pairs of program variables. Intuitively, $\vec{a}$ 
specifies which two variables should be tracked together by the analysis. 
Given such $\vec{a}$, the analysis defines the smallest 
partition $\Gamma$ of variables such that every $(x,y) \in \vec{a}$
is in the same partition of $\Gamma$. Then, it defines a grouped
Octagon domain~$\PODom = \prod_{\gamma \in \Gamma} \Oct_\gamma$ 
where $\Oct_\gamma$ is the usual Octagon domain for the variables in the partition
$\gamma$. The abstract domain of the analysis is $\mbc \to \PODom$, the collection
of maps from program points to grouped Octagons. The analysis performs fixed-point
computation on this domain using adjusted transfer functions of the
standard Octagon analysis. The details can be found in~\cite{HeoOhYa16}. 

We have to adjust the learning engine and our
feature-generation algorithm for this partial Octagon slightly. This is because the full Octagon
analysis on a program $P$ (that is, $\StaticAnalyzer(P,\vec{1})$) does not
work usually when the size of $P$ is large ($\ge$20KLOC in our experiments). Whenever the learning
part in Section~\ref{sec:learning} and our feature-generation algorithm have to run $\StaticAnalyzer(P,\vec{1})$ originally,
we run the impact pre-analysis in \cite{HeoOhYa16} instead. This pre-analysis is a fully relational
analysis that works on a simpler abstract domain than the full Octagon, and
estimates the behavior of the full Octagon; it defines a function $F^{\sharp} : \pgm \to \power{\query}$,
which takes a program and returns a set of queries in the program that are likely
to be verified by the full Octagon. Formally, we replaced
the predicate $\reducecond$ in Section~\ref{sec:generate-features} by
$\reducecond_q(P) = \left(q \not\in \StaticAnalyzer(P,\vec{0})  \myland q \in F^{\sharp}(P)\right)$.

\section{Experiments}
\label{sec:experiments}
% %\subsection{Setting}
% \label{sec:exp_setting}
% % We implemented the new techniques on top of Sparrow, a buffer-overrun
% % analysis that supports the full C language \cite{sparrow}.
% We evaluated our approach on benchmark programs from Linux and GNU
% packages. To automatically generate feature programs, we used C-Reduce,
% a general-purpose program reducer \cite{creduce}.

We evaluated our feature-generation algorithm with
the instance analyses in Section~\ref{sec:instances}.
We used the interval and Octagon analyses for proving the absence of buffer overruns,
and the pointer analysis the absence of null dereferences.

The three analyses are implemented on top
of our analysis framework for the C programming language~\cite{sparrow}.
The framework provides a baseline analysis that uses heap abstraction 
based on allocation sites and array smashing, is field-sensitive but context-insensitive, and performs
the sparse analysis~\cite{OhHLLY12,Farzan12,sparse2014,controlfeasibility-oopsla15}.
We extended this baseline analysis to implement the three analyses. Our pointer analysis uses 
Andersen's algorithm~\cite{Andersen94PhD}. 
The Octagon analysis is implemented by using the OptOctagons and Apron
libraries~\cite{Singh2015,apron}.
% For the decision tree learning, we used
% Scikit-learn~\cite{Pedregosa:2011tv}.
Our implementation of the feature-generation algorithm in
Section~\ref{sec:feat-generation} and the learning part in
Section~\ref{sec:learning} is shared by the three analyses except
that the analyses use slight variants of the $\dep$ function in
Section \ref{sec:learning}, which converts a query to program
components. In all three analyses, $\dep$ first computes a dependency
slice of a program for a given query.  Then, it collects program
variables in the slice for the interval and pointer analyses, and
pairs of all program variables in the slice for the Octagon
analysis. The computation of the dependency slice is approximate in
that it estimates dependency using a flow-insensitive pointer analysis
and ignores atomic commands too far away from the query when the size
of the slice goes beyond a threshold.\footnote{In our implementation,
  going beyond a threshold means having more than 200 program
  variables.}  This approximation ensures that the cost of computing
$\dep$ is significantly lower than that of the main analyses. We use
the same dependency analysis in $\fpred$. Our implementation uses
C-Reduce~\cite{creduce} for generating feature programs and a decision
tree algorithm~\cite{Pedregosa:2011tv} for learning a classifier for queries.
Our evaluation aims at answering four questions:
\begin{itemize}
\item \textbf{Effectiveness}: Does our feature-generation algorithm enable the learning of
        good analysis heuristics? % for each instance analysis? %Is it effective for various kinds of program analyses? 

\item \textbf{Comparison with manually-crafted features}:
 How does our approach of learning with automatically-generated features
        compare with the existing approaches of learning with
        manually-crafted features? 

\item \textbf{Impact of reducing and learning}:
 Does reducing a program really help for generating good features? Given a set
 of features, does learning lead to a classifier better than a simple (disjunctive)
 pattern matcher?

\item \textbf{Generated features}: Does our feature-generation algorithm produce
   informative features? %for each instance analysis?
  % Are the automatically generated
% features informative? Do they describe the key aspects of a given
% analysis task?

% \item \textbf{Benefits from Our Techniques}: Does the program
% reducer play an essential role in our approach? Does the automatic
% abstraction of data-flow graphs tailor the performance to particular
% analyses?

% \item \textbf{Alternative Approaches to Feature Learning}: Does our approach for automatic
%   feature generation have advantages
%   over alternative approaches? In
%   particular, how does it compare to using deep learning for
%   automatically learning features? 
\end{itemize}
% In our evaluation, we used codebases of open-source
% C programs collected from Linux and GNU packages,

\paragraph{Effectiveness}
\label{sec:effectiveness}

% The experimental results show that our method produces competitive
% analysis heuristics for three instance analyses.
% We demonstrate the effectiveness of our approach through partially
% flow-sensitive interval analysis, partial Octagon analysis,
% and partially flow-sensitive pointer analysis.

% \hy{I find it very strange that the split of programs into those for feature generation,
% the ones for learning, and the rest is different for each of the three analyses. This
% sounds like we manipulated the experimental setting. Also, in the main text, we say that
% the same set of programs is used both for feature generation and for learning. It looks
% like this is not the case. We should revise the main text.}

\begin{table*}
\small\
\centering
\begin{tabular}{|r|r|r|r|r|r|r|r|r|r|r|r|r|}
\hline
% 		  & \multicolumn{2}{c|}{Query Prediction} &
%                                                             \multicolumn{8}{c|}{Analysis}
% & \multicolumn{2}{c|}{Comparison}
% \\ \cline{2-13}
		  &  \multicolumn{2}{c|}{Query Prediction}  &
                                                              \multicolumn{3}{c|}{
                                                              \#Proved Queries} &
                                                       \multicolumn{3}{c|}{Analysis Cost (sec)}
                     & \multicolumn{2}{c|}{Quality}  & \multicolumn{2}{c|}{Oh et
                             al.~\cite{OhYaYi15}} \\ \cline{2-13}
	Trial & Precision & Recall & \multicolumn{1}{c|}{$\FIi$ ($a$)} &
                                                                       \multicolumn{1}{c|}{$\FSi$
                                                                       ($b$)}
                                                            &
                                                              \multicolumn{1}{c|}{
                                                              \Ours~($c$)}
                                                                                                    & \multicolumn{1}{c|}{$\FIi$ ($d$)} &
                                                                      \multicolumn{1}{c|}{$\FSi$
                                                                                                                                        }
                                                                                  &
                                                                                    \multicolumn{1}{c|}{\Ours~($e$)}
                                                                                    &
                                                                                      \multicolumn{1}{c|}{Prove}
              & \multicolumn{1}{c|}{Cost} & \multicolumn{1}{c|}{Prove} & \multicolumn{1}{c|}{ Cost} \\ \hline
	% === Copy-and-paste from Excel ===
1	&	71.5	\%	&	78.8	\%	&	6,537	&	7,126	&	7,019	&
										26.7	&	569.0	&	52.0	& 
										81.8 \%	&	1.9x	&
										56.6 \%	&	2.0x	\\
2	&	60.9	\%	&	75.1	\%	&	4,127	&	4,544	&	4,487	&
										58.3	&	654.2	&	79.9	&
										86.3 \%	&	1.4x	& 
										49.2 \%	&	2.4x	\\
3	&	78.2	\%	&	74.0	\%	&	6,701	&	7,532	&	7,337	&
										50.9	&	6,175.2	&	167.5	&
										76.5 \%	&	3.3x	& 
										51.1 \%	&	3.4x \\
4	&	72.9	\%	&	76.1	\%	&	4,399	&	4,956	&	4,859	&
										36.9	&	385.1	&	44.9	&
										82.6 \%	&	1.2x	& 
										54.8 \%	&	1.2x \\
5	&	83.2	\%	&	75.3	\%	&	5,676	&	6,277	&	6,140	&
										31.7	&	1,740.3	&	61.6	& 
										77.2 \%	&	1.9x	&
										65.6 \%	&	1.8x \\\hline
TOTAL	&	74.5	\%	&	75.7	\%	&	27,440	&	30,435	&	29,842	&
											204.9	&	9,523.9	&	406.1	&
											{\bf 80.2}	\%	&	{\bf 2.0x}	&
  											55.1 \%	&	2.3x\\\hline
	% =================================
\end{tabular}
\caption{Effectiveness of partially flow-sensitive interval
  analysis. Quality: $\mbox{Prove}=(c-a)/(b-a)$, $\mbox{Cost}=e/d$}
\label{tab:effectiveness-itv}
\vspace*{-3mm}
\end{table*}

 \begin{table*}
 \small
 \centering
 \begin{tabular}{|r|r|r|r|r|r|r|r|r|r|r|}
 \hline
  %       	  & \multicolumn{2}{c|}{Query Prediction} &	\multicolumn{8}{c|}{Analysis}	
  % \\ \cline{2-11}
 		  &  \multicolumn{2}{c|}{Query Prediction}  &
                                                              \multicolumn{3}{c|}{\#Proved Queries} &
                                                        \multicolumn{3}{c|}{Analysis Cost (sec)}
                      & \multicolumn{2}{c|}{Quality} \\ \cline{2-11}
 	Trial & Precision & Recall & \multicolumn{1}{c|}{$\FIp$} & \multicolumn{1}{c|}{$\FSp$} & \multicolumn{1}{c|}{ \Ours} & \multicolumn{1}{c|}{$\FIp$} &
                                                                       \multicolumn{1}{c|}{$\FSp$}
                                                                                   &
                                                                                     \multicolumn{1}{c|}{\Ours}
                                                                                     &
                                                                                       \multicolumn{1}{c|}{Prove}
               & \multicolumn{1}{c|}{Cost} \\ \hline
 	% === Copy-and-paste from Excel ===
 1	&	79.1	\%	&	76.8	\%	&	4,399
                   &	6,346	&	6,032	&	48.3	&
                                                                   3,705.0
                      &	150.0	&	83.9	\%	&	3.1x	 \\
 2	&	78.3	\%	&	77.1	\%	&	7,029
                   &	8,650	&	8,436	&	48.9	&
                                                                   651.4
                      &	74.0	&	86.8	\%	&	1.5x		\\
 3	&	74.5	\%	&	75.0	\%	&	8,781
                   &	10,352	&	10,000	&	41.5	&
                                                                   707.0
                      &	59.4	&	77.6	\%	&	1.4x		\\
 4	&	73.8	\%	&	75.9	\%	&	10,559
                   &	12,914	&	12,326	&	51.1	&
                                                                   4,107.0
                      &	164.3	&	75.0	\%	&	3.2x		\\
 5	&	78.0	\%	&	82.5	\%	&	4,205
                   &	5,705	&	5,482	&	23.0	&
                                                                   847.2
                      & 	56.7	&	85.1	\%	&	2.5x		\\\hline
 TOTAL	&	76.6	\%	&	77.3	\%	&	34,973
                      &	43,967	&	42,276	&	212.9	&	10,017.8	&	504.6
                      &	{\bf 81.2}	\%	&	{\bf 2.4x}		\\\hline
 	% =================================
 \end{tabular}
 \caption{Effectiveness of partially flow-sensitive pointer analysis}
 \label{tab:effectiveness-ptsto}
\vspace*{-3mm}
 \end{table*}

\begin{table*}
\small
\centering
\begin{tabular}{|r|r|r|r|r|r|r|r|r|r|r|r|r|}
\hline
		  % & \multicolumn{2}{c|}{Query Prediction} & \multicolumn{8}{c|}{Analysis} & \multicolumn{2}{c|}{Comparison} \\ \cline{2-13}
		  &  \multicolumn{2}{c|}{Query Prediction} &
                                                             \multicolumn{3}{c|}{
                                                             \#Proved
                                                             Queries}
                                                                                            &
                                                                                              \multicolumn{3}{c|}{Analysis
                                                                                              Cost
                                                                                              (sec)}
                  & \multicolumn{2}{c|}{Quality}&   \multicolumn{2}{c|}{Heo et al.~\cite{HeoOhYa16}} \\ \cline{2-13}
	Trial & Precision & Recall & \multicolumn{1}{c|}{$\FSi$} & \multicolumn{1}{c|}{$\Impact$} & \multicolumn{1}{c|}{\Ours} &
                                                                 \multicolumn{1}{c|}{$\FSi$}
                                                     & \multicolumn{1}{c|}{$\Impact$} &
                                                                   \multicolumn{1}{c|}{\Ours}
                                                                                  &
                                                                                    Prove
              & Cost & Prove & Cost \\ \hline
	% === Copy-and-paste from Excel ===
1	&	74.8	\%	&	81.3	\%	&	3,678	&	3,806	&	3,789	&	140.7	&	389.8	&	230.5	&	86.7	\%	&	1.6	x	&	100.0	\%	&	3.0	x	\\	
2	&	84.1	\%	&	82.6	\%	&	5,845	&	6,004	&	5,977	&	613.5	&	18,022.9	&	782.9	&	83.0	\%	&	1.3	x	&	94.3	\%	&	1.8	x	\\	
3	&	82.8	\%	&	73.0	\%	&	1,926	&	2,079	&	2,036	&	315.2	&	2,396.9	&	416.0	&	71.9	\%	&	1.3	x	&	92.2	\%	&	1.1	x	\\	
4	&	77.6	\%	&	85.2	\%	&	2,221	&	2,335	&	2,313	&	72.7	&	495.1	&	119.9	&	80.7	\%	&	1.6	x	&	100.0	\%	&	2.0	x	\\	
5	&	71.6	\%	&	78.4	\%	&	2,886	&	2,962	&	2,944	&	148.9	&	557.2	&	209.7	&	76.3	\%	&	1.4	x	&	96.1	\%	&	2.3	x	\\	\hline
TOTAL	&	79.0	\%	&	79.9	\%	&	16,556	&	17,186	&	17,067	&	1,291.0	&	21,861.9	&	1,759.0	&	\textbf{81.1}	\%	&	\textbf{1.4}	x	&	96.2	\%	&	1.8	x	\\	\hline
% =================================
\end{tabular}
\caption{Effectiveness of partial Octagon analysis}
\label{tab:effectiveness-oct}
\vspace*{-3mm}
\end{table*}

% \begin{table}[h]
% \small
% \centering
% \begin{tabular}{|r|c|c|c|c|}
% \hline
%       & \multicolumn{2}{c|}{Oh et al.~\cite{OhYaYi15}} & \multicolumn{2}{c|}{Ours} \\ \cline{2-5}
% Trial & Quality & Cost & Quality & Cost \\ \hline
% 1 & 85.2 \% & 1.5x & 88.7 \% & 1.4x \\
% 2 & 41.6 \% & 1.9x & 72.3 \% & 1.5x \\
% 3 & 89.9 \% & 3.2x & 84.0 \% & 6.4x \\
% 4 & 60.7 \% & 1.9x & 83.1 \% & 1.5x \\
% 5 & 47.8 \% & 2.7x & 94.2 \% & 1.7x \\ \hline
% TOTAL & 68.4 \% & 2.1x & 84.8 \% & 2.5x \\ \hline
% \end{tabular}
% \caption{Comparison with Oh et al.~\cite{OhYaYi15}}
% \label{tab:comparison-interval}
% \end{table}

% \begin{table}[h]
% \small
% \centering
% \begin{tabular}{|r|c|c|c|c|}
% \hline
%       & \multicolumn{2}{c|}{Heo et al.~\cite{HeoOhYa16}} & \multicolumn{2}{c|}{Ours} \\ \cline{2-5}
% Trial & Quality & Cost & Quality & Cost \\ \hline
% 1	&	100.0	\%	&	1.8	x	&	95.8	\%	&	1.9	x	\\	
% 2	&	100.0	\%	&	1.4	x	&	83.5	\%	&	1.2	x	\\	
% 3	&	85.4	\%	&	1.1	x	&	85.4	\%	&	1.5	x	\\	
% 4	&	100.0	\%	&	2.7	x	&	74.1	\%	&	2.8	x	\\	
% 5	&	100.0	\%	&	1.1	x	&	62.2	\%	&	2.6	x	\\	\hline
% Total	&	96.4	\%	&	1.7	x	&	82.7	\%	&	2.0	x	\\	\hline
% \end{tabular}
% \caption{Comparison with Heo et al.~\cite{HeoOhYa16}}
% \label{tab:comparison-octagon}
% \end{table}

%%% Local Variables:
%%% mode: latex
%%% TeX-master: "paper"
%%% End:

We compared the performance of our three instance analyses with their standard counterparts:
%in the following list:
\begin{itemize}
        \item Flow-insensitive($\FIi$) \& -sensitive($\FSi$) interval analyses
%          (in short, $\FIi$ and $\FSi$).
        \item Flow-insensitive($\FIp$) \& -sensitive($\FSp$) pointer analyses
        \item Flow-sensitive interval analysis ($\FSi$) and partial Octagon analysis 
                by impact pre-analysis ($\Impact$)~\cite{OhLHYY14}.
\end{itemize}
We did not include the Octagon analysis~\cite{octagon} in the list because the analysis
did not scale to medium-to-large programs in our benchmark set.

In experiments, we used 60 programs (ranging 0.4--109.6 KLOC) collected from
Linux and GNU packages. The programs are shown in Table \ref{benchmarks:flow-sensitivity}.
To evaluate the performance of learned heuristics
for the interval and pointer analyses, we randomly partitioned the 60 programs
into 42 training programs (for feature generation and learning) and 18 test programs
(for cross validation). For the Octagon analysis, we used only 25 programs out of 60
because for some programs, Octagon and the interval analysis prove the same set of queries.
We selected these 25 by running the impact pre-analysis~\cite{HeoOhYa16} for the Octagon on
all the 60 programs and choosing the ones that may benefit from Octagon according to the
results of this pre-analysis. The 25 programs are shown in Table \ref{benchmarks:octagon}.
We randomly partitioned the 25 programs into 17 training programs and 8 test programs.
From the training programs, we generated features and learned a heuristic based on these features.\footnote{We followed the practice used in representation learning~\cite{Bengio2013}, where both feature generation and learning are done with the same dataset.} The learned
heuristic was used for analyzing the test programs. We repeated this procedure for five times
with different partitions of the whole program sets. The average numbers of generated features
over the five trials were 38 (interval), 45 (pointer), and 44 (Octagon). C-Reduce took 0.5--24
minutes to generate a feature program from a query.All experiments were done on a Ubuntu machine
with Intel Xeon cpu (2.4GHz) and 192GB of memory.

% \hy{I find that the current organisation of the paragraphs is not very good because
% we sometimes say the same thing in each subsection. I was trying to reorganise the subsection.
% Assuming that we resolve the split issue, here is a plan to organise the rest of
% this subsection. Explain the split of programs into three groups, and other experimental
% setup. Describe the findings from the expriments. Finally, it is strange that from
% 15 programs we only get a small number of feature programs. If each program contains,
% say, 10 queries, we should get 150 feature programs. We need to explain why this blowup
% of the number of feature programs does not happen.}

% \hy{I won't revise the rest of the Experiments section this time. When it becomes 
% more complete later, I will have a go.}

% We compared the performance of our partially flow-sensitive interval
% analysis with a learned heuristic (Ours) against flow-insensitive (FI) and flow-sensitive (FS)
% analyses. The analyses prove the safety of each buffer access in a
% program.

%\paragraph{Partially Flow-Sensitive Analyses}

Table \ref{tab:effectiveness-itv} shows the performance of the learned
heuristics on the test
programs for the interval analysis. 
The learned classifier for queries (Section \ref{sec:learning})
was able to select $75.7\%$ of FS-provable but FI-unprovable queries on average
(i.e., $75.7\%$ recall) and $74.5\%$ of the selected queries were actually proved under
FS only (i.e., $74.5\%$ precision). With the analysis heuristic on top of this
classifier, our partially flow-sensitive analysis could prove $80.2\%$
of queries that require flow-sensitivity while increasing the time of the flow-\emph{in}sensitive analysis by $2.0$x on average.
% The time for pre-processing a program (i.e., generating data dependencies and matching
% features against queries) is included in the result. 
 We got $80.2$ (higher than $75.7$) because
the analysis parameter is the set of all the program components for queries
selected by the classifier and this parameter may make the analysis prove queries
not selected by the classifier.
The fully flow-sensitive analysis increased the analysis time by
$46.5$x. 
We got similar results for the other two analyses (Tables~\ref{tab:effectiveness-ptsto} 
and \ref{tab:effectiveness-oct}).

% The learned heuristics generalized well on unseen programs.  The last
% column of Table \ref{tab:effectiveness-itv} (Self) shows the performance of
% the heuristics learned with the test programs;
% the analysis with these self-trained (or overfit) heuristics proved, on
% average, 84.6\% of provable queries, which is similar to the
% performance (84.8\%)
% obtained by training the heuristics with the training, not test, programs.

% We got similar results for the other two analyses (Tables~\ref{tab:effectiveness-ptsto} 
% and \ref{tab:effectiveness-oct}). Overall, these two analyses proved
% $81.9\%$ and $79.8\%$ of the queries that the most accurate analysis
% or its impact pre-analysis can prove. They increased the analysis cost
% of the most inaccurate counterparts by $2.5$x and $1.4$x only.
% The most accurate analyses, $\FSp$ and $\Impact$, increased the
% analysis cost by 48x and 17x, respectively. 

\paragraph{Comparison with Manually-Crafted Features}

We compared our approach with those in Oh et al.~\cite{OhYaYi15} and
Heo et al.~\cite{HeoOhYa16}, which learn
 analysis heuristics using manually-crafted
 features. The last two columns of Table~\ref{tab:effectiveness-itv}
 present the performance of the partially
 flow-sensitive interval analysis in~\cite{OhYaYi15},
 and those of Table~\ref{tab:effectiveness-oct} the
 performance of the partial Octagon analysis in~\cite{HeoOhYa16}. 

The five
trials in the tables use the splits of training and test programs in the corresponding entries
of Tables~\ref{tab:effectiveness-itv} and~\ref{tab:effectiveness-oct}. Our implementation of 
Oh et al.'s approach used their 45 manually-crafted features, and
applied their Bayesian optimization algorithm to our benchmark programs.
Their approach requires the choice of a threshold value $k$, which determines
how many variables should be treated flow-sensitively. For each trial and each program in that trial, 
we set $k$ to the number of variables selected by our approach, so that both approaches induce
similar overhead in analysis time. Our implementation of Heo et al.'s
approach used their 30 manually-crafted features, and applied their
supervised learning algorithm to our benchmark programs. 

The results show that our approach is on a par with the existing
ones, while not requiring the manual feature design.  For the
interval analysis, our approach consistently proved more queries than
Oh et al.'s ($80.2\%$ vs $55.1\%$ on average).  For Octagon,
Heo et al.'s approach proved more queries than ours ($81.1\%$ vs
$96.2\%$). %but increased the cost on average.
We warn a reader that these are just end-to-end comparisons and it is
difficult to draw a clear conclusion, as the learning algorithms of
the three approaches are different. However, the overall results show
that using automatically-generated features is as competitive as using
features crafted manually by analysis designers.

% One interesting observation is that learned heuristics by our
% approach tend to generalize well to unseen testing
% programs. In~\cite{OhYaYi15}, a learned heuristic proved 84.0\% of
% provable queries on training programs but 69.6\% on test
% programs. We observed similar overfitting with our benchmark
% programs as well. Our conjecture is that learning with
% manually-crafted features is likely to overfit because feature
% engineering is expensive and often done by investigating only a
% small set of training programs and queries. On the other hand, our
% approach generates features from a wide range of programs and
% queries, and may avoid such overfitting, as shown in our
% experiments.  d \hy{Why don't we report experimental results for the
% other two analyses?}

\paragraph{Impact of Reducing and Learning}

In order to see the role of a reducer in our approach, we generated
feature programs without calling the reducer in our experiment with
the interval analysis. These unreduced feature programs were then
converted to abstract data-flow graphs or features, which enabled the
learning of a classifier for queries.  The generated features were too
specific to training programs, and the learned classifier did not
generalize well to unseen test programs; removing a
reducer dropped the average recall of the classifier from
$75.7\%$ to $58.2\%$ for test programs.

In our approach, a feature is a reduced and abstracted program slice
that illustrates when high precision of an analysis is useful for
proving a query. Thus, one natural approach is to use the disjunction
of all features as a classifier for queries. Intuitively, this
classifier attempts to pattern-match each of these features against a
given program and a query, and it returns true if some attempt
succeeds. We ran our experiment on the interval analysis with
this disjunctive classifier, instead of the original decision tree
learned from training programs. This change of the classifier
increased the recall from $75.7\%$ to $79.6\%$, but
dropped the precision significantly from $74.5\%$ to
$10.4\%$. The result shows the benefit of going beyond the simple
disjunction of features and using a more sophisticated boolean
combination of them (as encoded by a decision tree). One possible
explanation is that the matching of multiple features suggests 
the high complexity of a given program, which typically makes the analysis
lose much information even under the high-precision setting.

\paragraph{Generated Features}
\label{sec:feature_discussion}

We ranked generated features in our experiments according to their Gini scores~\cite{Breiman:2001fb} in the learned 
decision tree, which measure the importance of these features for
prediction. 
For each instance analysis, we show 
two features that rank consistently high in the five trials of 
experiments.  For readability, we present them
in terms of their feature programs, rather than as abstract data-flow
graphs.

The top-two feature programs for the interval analysis and the pointer analysis are:

\begin{Verbatim}
// Feature Program 1 for Interval
int buf[10];
for (i=0;i<7;i++) { buf[i]=0; /* Query */ }
\end{Verbatim}
\begin{Verbatim}
// Feature Program 2 for Interval
i=255; p=malloc(i);
while (i>0) { *(p+i)=0; /* Query */  i--; }
\end{Verbatim}
\begin{Verbatim}
// Feature Program 1 for Pointer
i=128; p=malloc(i); 
if (p==0) return; else *p=0; /* Query */
\end{Verbatim}
\begin{Verbatim}
// Feature Program 2 for Pointer
p=malloc(i); p=&a; *p=0; /* Query */
\end{Verbatim}

\noindent
The feature programs for the interval analysis describe cases where a
consecutive memory region is accessed in a loop through increasing or
decreasing indices and these indices are bounded by a constant from
above or from below because of a loop condition. The programs for the
pointer analysis capture cases where the safety of pointer dereference
is guaranteed by null check or preceding strong update. All of these
programs are typical showcases of flow-sensitivity.
% : in those programs,
% flow-sensitive information is needed for proving queries and at the
% same time this information is easy enough to be inferred automatically
% by analyses.
%But we found it interesting that they form the predictive features in the learned classifiers.

% \hy{We say that we tried the experiment with each analysis five times. In each trial,
% the set of found features can be different. It is not clear to me which trial is used
% for finding these top-two features. Also, one might ask whether the two features above rank
% consistently high in all five trials. Otherwise, their high ranks might come from random noise.}

The top-two feature programs for the partial Octagon analysis are: 
\begin{Verbatim}
// Feature Program 1 for Octagon        
size=POS_NUM; arr=malloc(size); 
arr[size-1]=0; /* Query */
\end{Verbatim}
\begin{Verbatim}
// Feature Program 2 for Octagon        
size=POS_NUM; arr=malloc(size);
for(i=0;i<size;i++){ arr[i]=0; /* Query */ }
\end{Verbatim}
These feature programs allocate an array of a positive size
and access the array using an index that is related to this size
in a simple linear way. They correspond to our expectation about
when the Octagon analysis is more effective than the interval analysis.

When converting feature programs to abstract data-flow graphs, 
our approach automatically identifies the right abstraction level of
commands for each instance analysis (Algorithm \ref{alg:overall}). 
In the interval analysis, the abstraction found by our approach was to merge
all the comparison operators (e.g., $<,\le,>,\ge,=$) but to differentiate
all the arithmetic operators (e.g., $+$, $-$, $*$).
This is because, in the interval analysis, comparison with constants is generally
a good signal for improving precision regardless of a specific comparison operator used,
but the analysis behaves differently when analyzing different commands involving $+$ or $-$.
With other abstractions, we obtained inferior performance; for
example, when we
differentiate comparison operators and abstract away arithmetic
operators, the recall was dropped
from $75.7\%$ to $54.5\%$. % \kwonsoo{Fixed the precision and recall here. Need better interpretation?}
%the precision and recall of the learned
%classifier were dropped to $79.7\%$ and $47.0\%$, respectively. 
In the pointer analysis, the best abstraction was to
merge all arithmetic and comparison operators while differentiating
equality operators ($=, \neq$).
% \kwonsoo{Is it okay to write $\neq$? This operator does not appear in
% Figure 2.}
% \kwonsoo{In the pointer analysis, differentiating all arithmetic and comparison operators
% (e.g., $+$, $-$, $*$, $<,\le,>,\ge, =$) led to the highest score (i.e., $\hat{R}=\{\}$).
% Merging all comparison operators ($<,\le,>,\ge$, =, !=) while differentiating
% arithmetic operators (e.g., $+$, $-$, $*$) led to the lowest score.
% The differences are marginal, though. The highest and lowest F1-score was
% 80.27 and 77.76, respectively.}
% \kwonsoo{In current implementation, we somehow abstract other program parts.
% For example, 'match' returns true with *var and (var + \_). This abstraction cannot
% be found by our algorithm, because they do not share the same LHS in the grammar rules. Also, finding them automatically is nearly infeasible in general. Probably other abstractions that
% our algorithm cannot identify contribute to bigger differences.
% It might be just an implementation detail, but I wrote it here for your information.}
In the Octagon analysis, our approach identified an 
abstraction that merges all comparison and binary operators while
differentiating addition/subtraction operators from them. 

% \hy{Check whether we want to say something about pointer analysis. In the original write-up,
% Kihong mentioned ``linear relationships in a program''. I removed that phrase,
% because I couldn't understand what he meant. Check whether the editted text says something correct.}

\longversion{
\paragraph{Limitations}

Our current implementation has two 
limitations. First, because we approximately generate data dependency (in $\dep$ and $\fpred$)
within a threshold, we cannot apply our approach to instances
that require computing long dependency chains, e.g., 
context-sensitive analysis. 
%Although this is not a
%fundamental limitation, 
We found that computing dependency slices of
queries beyond procedure boundaries efficiently with enough 
precision is hard to achieve in practice.  Second, our method could not be
applicable to program analyses for other programming languages (e.g.,
oop, functional), as we assume that a powerful program reducer and a
way to efficiently build data-flow graphs exist for the target
language, which does not hold for, e.g., JavaScript. 
}

\section{Related Work}
\label{sec:related}

\paragraph{Parametric Static Analysis}

% A parametric static analysis % comes with
% % a class of program abstractions and 
% tries to find an appropriate
% abstraction for a given program. % , which balances the cost and the 
% % precision of analyzing the program.
In the past decade, a large amount of research in static analysis has
been devoted for developing an effective heuristic for finding a good
abstraction. Several effective techniques based on
counterexample-guided abstraction refinement
(CEGAR)~\cite{slam,clarke:jacm03,blast:popl04,rybal1,gulavani:tacas08,XinMGNY14,XinNY13}
have been developed, which iteratively refine the abstraction based on
the feedback from the previous runs.
% search for a good abstraction % for a given
% % program
% by repeatedly analyzing the program under different abstractions and
% choosing the next abstraction based on the feedback from the previous
% analysis runs.
Other techniques choose an abstraction using dynamic analyses~\cite{NaikYCS12,GuptaMR13} or
pre-analyses~\cite{Smaragdakis:2014,OhLHYY14}. Or they simply use a
good manually-designed heuristic, such as the one for controlling the
object sensitivity for Java programs~\cite{Kastrinis2013}. In all of
these techniques, heuristics for choosing an abstraction cannot
automatically extract information from one group of programs, and
generalize and apply it to another group of programs. This
cross-program generalization in those techniques is, in a sense, done
manually by analysis designers.

Recently, researchers have proposed new techniques for finding
effective heuristics automatically rather than
manually~\cite{Grigore2016,OhYaYi15,HeoOhYa16,ChaJeOh16}.  In these
techniques, heuristics themselves are parameterized by
hyperparameters, and an effective hyperparameter is learned from
existing codebases by machine learning algorithms, such as Bayesian
optimization and decision tree
learning~\cite{OhYaYi15,HeoOhYa16,ChaJeOh16}.
% These techniques have been applied to learn good heuristics (i.e., hyperparameters) for
% controlling flow-sensitivity~\cite{OhYaYi15}, context-sensitivity~\cite{OhYaYi15}, the
% degree of relational analysis~\cite{HeoOhYa16}, and the thresholds of widening~\cite{ChaJeOh16}. 
In~\cite{Grigore2016}, a learned hyperparameter determines a
probabilistic model, which is then used
to guide an abstraction-refinement algorithm. % based on Datalog.

Our work improves these recent techniques by addressing the important issue of feature design.
The current learning-based techniques assume well-designed features, and leave the obligation of
discharging this nontrivial assumption to analysis designers~\cite{OhYaYi15,ChaJeOh16,HeoOhYa16}.
The only exception % that we are know of
is \cite{Grigore2016}, but the technique there
applies only to a specific class of program analyses written in Datalog. In particular,
it does not apply to the analyses with infinite abstract
domains, such as interval and Octagon analyses. %considered in this paper. 
Our work provides a new automatic way of discharging the assumption on feature design, which can be applicable 
to a wide class of program analyses because our approach uses program analysis as a black box
and generates features (i.e., abstracted small programs) not tied to the internals of the analysis.

\paragraph{Application of Machine Learning in Program Analysis}

Recently machine learning techniques have been applied for addressing
challenging problems in program analysis. They
have been used for generating candidate invariants from data collected from 
testing~\cite{Garg2014,Garg2016,Nori2013,Sharma2013,Sharma2012}, 
for discovering intended behaviors of programs
(e.g., preconditions of functions, API usages, types, and
informative variable names)~\cite{Gehr2015,Padhi2016,MishneSY12,BeckmanN11,LivshitsNRB09,Raychev2015,Sankaranarayanan2008,Zhu2016,Zhu2015,Allamanis2015,Katz2016,Kulkarni2016},
for finding semantically similar pieces of code~\cite{David2016}, 
and for synthesizing programs 
(e.g., code completion and patch generation)~\cite{Raychev2014,Hindle2012,Long2016,Allamanis2016,Beilik2016,RaychevBVK16}.
Note that the problems solved by these applications are different from ours, which
is to learn good analysis heuristics from existing codebases and to generate good 
features that enable such learning.

\paragraph{Feature Learning in Machine Learning}

Our work can be seen as a feature learning technique specialized to
the program-analysis application. Automating the feature-design
process has been one of the holy grails in the machine learning
community, and a large body of research has been % devoted for this
% issue
done
under the name of representation learning or feature
learning~\cite{Bengio2013}. Deep learning~\cite{DeepLearning} is
perhaps the most successful feature-learning
method, which simultaneously learns features and classifiers
 through multiple layers of representations. It has been recently applied to programming tasks
(e.g.~\cite{Allamanis2016}). A natural question is, thus, whether deep
learning can be used to learn program analysis heuristics 
as well.  % Our preliminary experience is that learning program analysis
% heuristics from raw data (i.e., program text) via deep learning has
% potential but requires engineering efforts and expertise in deep
% learning. Specifically,
In fact, we trained a character-level convolutional
network in Zhang et al.~\cite{Zhang2016Nips} for predicting % whether a
% flow-sensitive interval analysis proves a given query or not.
the need for flow-sensitivity in interval analysis. 
We
represented each query by the 300 characters around the query in the
program text, and used pairs of character-represented query and its
provability as training data. We tried a variety of settings (varying,
e.g., \#layers, width, \#kernels, kernel size, activation functions,
\#output units, etc) of the network, but the best performance we could
achieve was 93\% of recall with disappointing 27\% of precision on
test data. Achieving these numbers was highly nontrivial, and we could
not find intuitive explanation about why a particular setting of the
network leads to better results than others. We think that going
beyond 93\% recall and 27\% precision in this application is challenging 
and requires expertise in deep learning.

\section{Conclusion}
\label{sec:conclusion}

We have presented an algorithm that mines existing codebases and
generates features for a data-driven parametric 
static analysis. The generated features enable the learning of an
analysis heuristic from the codebases, which decides whether each
part of a given program should be analyzed under high precision 
or under low precision.
The key ideas behind the algorithm are to use abstracted code snippets
as features, and to generate such snippets using a program reducer.
We applied the algorithm to partially flow-sensitive interval and pointer
analyses and partial Octagon analysis. Our experiments with these analyses
and 60 programs from Linux and GNU packages 
show that the learned heuristics with automatically-generated features achieve
performance comparable to those with manually-crafted features. 

Designing a good set of features is a nontrivial and costly step in
most applications of machine learning techniques. % , and has been the subject
% of a large amount of research in the machine learning community.
We hope
that our algorithm for automating this feature design for data-driven program
analyses or its key ideas help attack this feature-design problem in the
ever-growing applications of machine learning for program analysis, verification,
and other programming tasks.

%%% Local Variables:
%%% mode: latex
%%% TeX-master: "paper"
%%% End:

\bibliographystyle{plain}
\bibliography{refs}

\appendix
\section{Benchmark Programs}

Table \ref{benchmarks:flow-sensitivity} and \ref{benchmarks:octagon}
show the benchmark programs for the partially flow-sensitive interval
and pointer analyses, and the partial Octagon analysis, respectively. 

% In experiments, we used 60 programs collected from Linux and GNU
% software packages. The list of 60 programs is shown in Table
% \ref{benchmarks:flow-sensitivity}. Those programs were used to
% evaluate our partially flow-sensitive interval and pointer
% analyses. For the partial Octagon analysis, we used 25 programs out of
% 60, which are shown in Table.

\begin{table*}[t]
\centering
\small
\begin{tabular}{|l|r|l|r|}
\hline
Programs & LOC & Programs & LOC \\ \hline
brutefir-1.0f	&	398	&	mpage-2.5.6	&	14,827	\\	
consol\_calculator	&	1,124	&	bc-1.06	&	16,528	\\	
dtmfdial-0.2+1	&	1,440	&	ample-0.5.7	&	17,098	\\	
id3-0.15	&	1,652	&	irmp3-ncurses-0.5.3.1	&	17,195	\\	
polymorph-0.4.0	&	1,764	&	tnef-1.4.6	&	18,172	\\	
unhtml-2.3.9	&	2,057	&	ecasound2.2-2.7.0	&	18,236	\\	
spell-1.0	&	2,284	&	gzip-1.2.4a	&	18,364	\\	
mp3rename-0.6	&	2,466	&	unrtf-0.19.3	&	19,019	\\	
mp3wrap-0.5	&	2,752	&	jwhois-3.0.1	&	19,375	\\	
ncompress-4.2.4	&	2,840	&	archimedes	&	19,552	\\	
pgdbf-0.5.0	&	3,135	&	aewan-1.0.01	&	28,667	\\	
mcf-spec2000	&	3,407	&	tar-1.13	&	30,154	\\	
acpi-1.4	&	3,814	&	normalize-audio-0.7.7	&	30,984	\\	
unsort-1.1.2	&	4,290	&	less-382	&	31,623	\\	
checkmp3-1.98	&	4,450	&	tmndec-3.2.0	&	31,890	\\	
cam-1.05	&	5,459	&	gbsplay-0.0.91	&	34,002	\\	
bottlerocket-0.05b3	&	5,509	&	flake-0.11	&	35,951	\\	
129.compress	&	6,078	&	enscript-1.6.5	&	38,787	\\	
e2ps-4.34	&	6,222	&	twolame-0.3.12	&	48,223	\\	
httptunnel-3.3	&	7,472	&	mp3c-0.29	&	52,620	\\	
mpegdemux-0.1.3	&	7,783	&	bison-2.4	&	59,955	\\	
barcode-0.96	&	7,901	&	tree-puzzle-5.2	&	62,302	\\	
stripcc-0.2.0	&	8,914	&	icecast-server-1.3.12	&	68,571	\\	
xfpt-0.07	&	9,089	&	dico-2.0	&	69,308	\\	
man-1.5h1	&	11,059	&	aalib-1.4p5	&	73,413	\\	
cjet-0.8.9	&	11,287	&	pies-1.2	&	84,649	\\	
admesh-0.95	&	11,439	&	rnv-1.7.10	&	93,858	\\	
hspell-1.0	&	11,520	&	mpg123-1.12.1	&	101,701	\\	
juke-0.7	&	12,518	&	raptor-1.4.21	&	109,053	\\	
gzip-spec2000	&	12,980	&	lsh-2.0.4	&	109,617	\\	\hline
\end{tabular}
\caption{60 benchmark programs for our partially flow-sensitive
  interval and pointer analyses. 
}
\label{benchmarks:flow-sensitivity}
\end{table*}

\begin{table*}
\centering
\small
\begin{tabular}{|l|r|l|r|}
\hline
Programs & LOC & Programs & LOC \\ \hline
brutefir-1.0f	&	398	&	ecasound2.2-2.7.0	&	18,236	\\	
consol\_calculator	&	1,124	&	unrtf-0.19.3	&	19,019	\\	
dtmfdial-0.2+1	&	1,440	&	jwhois-3.0.1	&	19,375	\\	
id3-0.15	&	1,652	&	less-382	&	31,623	\\	
spell-1.0	&	2,284	&	flake-0.11	&	35,951	\\	
mp3rename-0.6	&	2,466	&	mp3c-0.29	&	52,620	\\	
e2ps-4.34	&	6,222	&	bison-2.4	&	59,955	\\	
httptunnel-3.3	&	7,472	&	icecast-server-1.3.12	&	68,571	\\	
mpegdemux-0.1.3	&	7,783	&	dico-2.0	&	69,308	\\	
barcode-0.96	&	7,901	&	pies-1.2	&	84,649	\\	
juke-0.7	&	12,518	&	raptor-1.4.21	&	109,053	\\	
bc-1.06	&	16,528	&	lsh-2.0.4	&	109,617	\\	
irmp3-ncurses-0.5.3.1	&	17,195	&		&		\\	\hline
\end{tabular}
\caption{25 benchmark programs for 
  our partial Octagon analysis. }
\label{benchmarks:octagon}
\end{table*}

%%% Local Variables:
%%% mode: latex
%%% TeX-master: "supplementary"
%%% End:

%\input{issue}

\end{document}